\documentclass[twocolumn]{aastex631}
\begin{document}

\newcommand{\au}{{\rm{AU}}}
\newcommand{\rebound}{\texttt{REBOUND}}
\newcommand{\mercurius}{\texttt{MERCURIUS}}
\newcommand{\unitacc}{{{\rm cm\ s^{-2}}}}
\newcommand{\gcc}{{{\rm g\ cm^{-3}}}}
\newcommand{\rhopl}{{{\rho_{\rm{pl}}}}}
\newcommand{\mpl}{{{m_{\rm{pl}}}}}
\newcommand{\Ppl}{{{P_{\rm{pl}}}}}
\newcommand{\mdisk}{{{m_{\rm{d}}}}}
\newcommand{\Npl}{{{N_{\rm{pl}}}}}
\newcommand{\mplanet}{{{m_{\rm{p}}}}}
\newcommand{\tramp}{{{t_{\rm{ramp}}}}}
\newcommand{\nres}{{N_{\rm{res}}}}
\newcommand{\nnonres}{{N_{\rm{non-res}}}}
\newcommand{\emcee}{{{\texttt emcee}}}
\newcommand{\epsilonobs}{{\epsilon_{\rm{obs}}}}
\newcommand{\fracdelm}{{\Delta\mplanet/\mplanet}}
\newcommand{\epsilonini}{{\epsilon_{\rm{ini}}}}
\newcommand{\epsilonfin}{{\epsilon_{\rm{fin}}}}
\newcommand{\porb}{{P_{\rm{orb}}}}
\newcommand{\nobs}{{N_{\rm{obs}}}}
\newcommand{\ntot}{{N_{\rm{total}}}}
\newcommand{\minteract}{{m_{\rm{int}}}}
\newcommand{\imut}{{i_{\rm{m}}}}

\title{Effects of Planetesimal Scattering:\\
Explaining the Observed Offsets from Period Ratios 3:2 and 2:1}

\author[0000-0002-3103-2000]{Tuhin Ghosh}
\affiliation{Department of Astronomy and Astrophysics, Tata Institute of Fundamental Research, Homi Bhabha Road, Navy Nagar, Colaba, Mumbai, 400005, India}
\email{tghosh.astro@gmail.com}

\author[0000-0002-3680-2684]{Sourav Chatterjee}
\affiliation{Department of Astronomy and Astrophysics, Tata Institute of Fundamental Research, Homi Bhabha Road, Navy Nagar, Colaba, Mumbai, 400005, India}
\email{souravchatterjee.tifr@gmail.com}

\shorttitle{Effects of Planetesimal Scattering}
\shortauthors{Ghosh \& Chatterjee}

\begin{abstract}
The observed deficit and excess of adjacent planet pairs with period ratios narrow and wide of $3:2$ and $2:1$, the nominal values for the corresponding mean motion resonances (MMRs), have intrigued many. Previously, using a suite of simulations, \citet{Chatterjee&Ford2015} showed that the excess above the $2:1$ MMR can be naturally explained if planet pairs, initially trapped in the $2:1$ MMR, dynamically interact with nearby planetesimals in a disk. We build on this work by: a) updating the census of discovered planet pairs, b) extending the study to initially non-resonant as well as resonant planet pairs, c) using initial planet and orbital properties directly guided by those observed, and d) extending the initial period ratios to include both $2:1$ and $3:2$. We find that 1) interactions with planetesimals typically increase the period ratios of both initially resonant and non-resonant planet pairs; 2) starting from an initially flat period ratio distribution for systems across $3:2$ and $2:1$, these interactions can naturally create the deficits observed narrow of these period ratios; 3) contribution from initially resonant planet pairs is needed to explain the observed levels of excess wide of $3:2$; 4) a mixture model where about $25\%$ ($1\%$) planet pairs were initially trapped into $3:2$ ($2:1$) MMRs is favored to explain both the observed deficit and excess of systems across these period ratios, however, up to a few percent of planet pairs are expected to remain in MMR today.    
\end{abstract}

\section{Introduction} \label{sec:intro}

The Nobel-winning discovery of 51 Pegasi b, the first planet discovered around a main-sequence star (other than the Sun), by \citet{1995Natur.378..355M} has since started the most exciting field in modern astrophysics. Thanks to numerous successful space-based missions such as NASA's {\em Kepler} 
\citep{2010Borucki, Borucki_2016}, {\em K2} \citep{Howell_2014, Cleve_2016}, and {\em TESS} \citep{2015RickerTESS}, and numerous radial velocity surveys such as HIRES \citep{HIRES}, HARPS \citep{HARPS}, HARPS-N \citep{HARPS-N}, SOPHIE \citep{SOPHIE}, and ESPRESSO \citep{ESPRESSO} and follow-up missions like LAMOST-{\em Kepler} survey \citep{Lamost_Dong_2014, Lamost_Luo_2015} and California-{\em Kepler} Survey \citep{CKS}, we now know of more than five thousand exoplanets \citep[NASA Exoplanet Archive, \url{http://exoplanetarchive.ipac.caltech.edu}][]{2013Akeson_NEA} in more than $\sim3500$ planetary systems. The large number of discovered planetary systems not only tell us about the unprecedented diversity of present-day orbital and structural properties of exoplanetary systems, they also provide clues towards our still evolving understanding of the formation and dynamical history of these planetary systems \citep[e.g.,][]{2011_Ford, 2011Lissauer, 2012_Fang, 2012Rein, 2013_Hansen_Murray, 2014Fabrycky, 2015_Malhotra, 2015_Pu_Wu, 2015Steffen, 2016_Ballard_Johnson}.

\begin{figure}[htb!]
\plotone{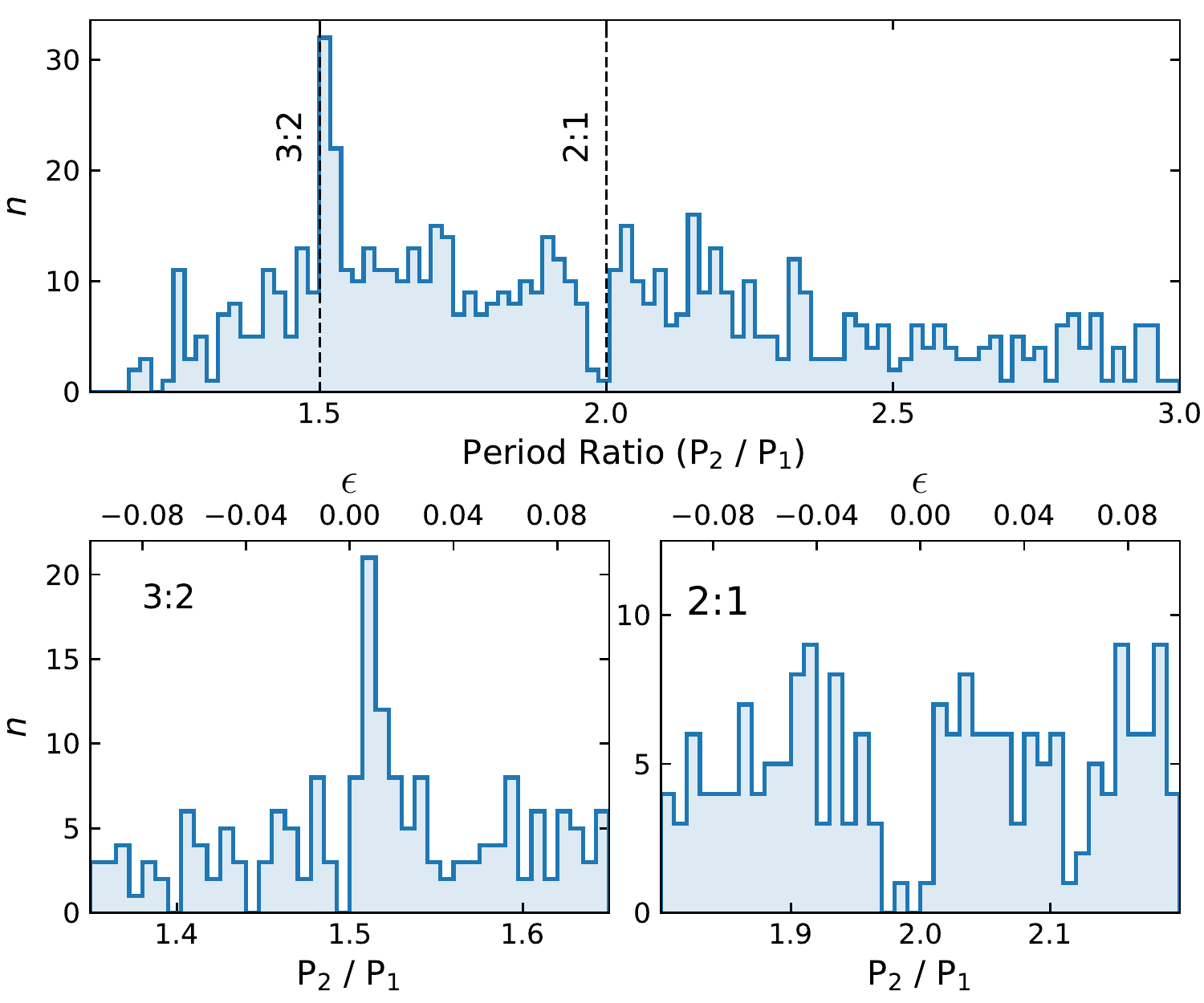}
\caption{{\em Top}: Histogram showing the period ratio distribution of all adjacent planet pairs smaller than Neptune. Data is taken from the NASA Exoplanet Archive on August 12, 2022. The vertical dotted lines denote the nominal positions of the $3:2$ and $2:1$ MMRs. {\em Bottom}: Same as the top panel but zoomed in around the MMRs. The secondary horizontal axis shows $\epsilon$, the offset from the nominal MMR positions ($\epsilon=0$). 
\label{fig:pr_obs}}
\end{figure}
One of the most curious observed trends among multi-planet systems is the overabundance and deficit of systems of adjacent planet pairs just wide and narrow of two major first-order mean motion resonances $3:2$ and $2:1$ with large offsets $\epsilon$ \citep{2011Lissauer, 2014Fabrycky}. Throughout this study we use $\epsilon\equiv(P_2/P_1)/[(p+1)/p] -1$ as a measure of the offset from the nominal position of the $(p+1):p$ MMR. \autoref{fig:pr_obs} shows the distribution of observed period ratios for adjacent planet pairs. The excess of systems wide of $3:2$ and deficit narrow of $2:1$ are quite prominent. Indeed, \citet{2015Steffen} showed that in particular, the deficit narrow of the $2:1$ MMR and the excess wide of the $3:2$ MMR, are  statistically significant to $\approx 99\%$ confidence level, while, the excess wide of $2:1$ and deficit narrow of $3:2$ are not as significant. When first discovered, this was particularly surprising since overall, the period ratios of adjacent planet pairs seem to show little preference for the nominal resonance positions in contrast to earlier findings involving jovian planet pairs that exhibit a clear preference for the $2:1$ period commensurability with small offsets \citep[e.g.,][]{Butler2006, 2011_Wright}. This abundance asymmetry across these period ratios seems to indicate that the near-resonance planet pairs know the existence of these MMRs, but somehow manage to avoid the nominal period ratios with large positive offsets.

This discovery generated widespread interest since the existence or absence of MMRs among planet pairs can shed light on the formation of planetary systems and their subsequent evolution. For example, on one hand, the predominance of resonant planet pairs indicates convergent gas-disk-driven migration and efficient trapping in stable resonances with little or insufficient subsequent perturbations \citep[e.g.,][]{Goldreich1980,2002LeePeale,Thommes2003,Kley2012}. On the other hand, the absence of two- (or more-) planet MMRs can put constraints either on the success rate of resonance trapping due to convergent migration \citep[e.g.,][]{Goldreich_2014, Batygin_2015, Deck2015} or the importance of subsequent dynamical processes that effectively break MMR configurations \citep[e.g.,][]{Chatterjee_2008,Matsumura2010,2017Izidoro, 2018Ogihara}.

Several theoretical investigations have proposed a variety of mechanisms to explain this curious observed trend. However, a general consensus remains illusive. For example, whether most of the near-resonant systems are in resonance or not is yet unknown. It was pointed out that eccentricity damping of planetary orbits while trapped in an MMR may create large offsets from the nominal commensurate period ratios \citep[e.g.,][]{Lithwick_2012}. The mechanism for the eccentricity damping, however, created significant debate. Several studies proposed that star-planet tides may be responsible for eccentricity damping at levels necessary for creating the observed trend \citep[e.g.,][]{2013Batygin, 2014Delisle_Laskar, Xie_2014}. In contrast, other studies pointed out that for the star-planet tides to be solely responsible for the observed trend, tides either need to be unusually strong \citep[e.g.,][]{Lithwick_2012} or perhaps there is another source of damping \citep[e.g.,][]{Lee_2013,2015_Silburt_Rein}. It was also pointed out that the asymmetry does not seem to be particularly sensitive to the distance from the respective host stars, as would be expected if star-planet tidal damping were fully responsible for the observed asymmetry \citep[e.g.,][]{2020Choksi}. Other studies have suggested that the observed asymmetry can be attributed to the details of the disk-driven migration process itself. For example, \citet{2017Ramos} have suggested that the distribution of near-resonant planets in the super-earth mass range may be dependent on the details of the disk properties including how flared the disk is. On the other hand, \citet{2020Choksi} and \citet{2021Wang} have suggested that in presence of a gas disk, the migrating planets simply get trapped wide of the nominal period ratios due to continued eccentricity damping by the gas disk. Analyzing a sample of three-planet systems, where each pair has a period ratio just wide of the nominal MMR, \citet{2019_Pichierri} have concluded that some of these triplets could be locked in multi-resonant chains trapped via convergent migration, while subsequent gas-disk dissipation repels the orbits wide of the exact commensurabilities. On the contrary, \citet{Veras2012} have suggested that a high fraction of the observed near-resonance planet pairs may not truly be in MMR.

While smooth gas-disk driven migration is expected to trap planets in MMRs, stochastic processes such as dynamical scattering have the potential for breaking the MMRs. Motivated by this expectation, several studies combined smooth migration with some source of stochasticity, for example, from turbulent clumps in the gas disk \citep[e.g.,][]{2012Rein,2017Batygin} or due to interactions with a residual disk of planetesimals after gas disk is sufficiently depleted \citep[][hereafter CF15]{Chatterjee&Ford2015}. On the other hand, \citet{Petrovich_2013} suggested that the observed asymmetry may be a result of the planet formation process itself by showing in idealised cases that as planets grow, test particles tend to pile up wide of the first-order MMRs with respect to the growing planet.

Compact multiplanet systems are also susceptible to dynamical instability even without any external perturbations \citep[e.g.,][]{1996Chambers}. In non-resonant systems, planet pairs close to MMRs tend to get destabilized on shorter timescales than pairs away from MMRs \citep[e.g.,][]{Chatterjee_2008,2010Funk, 2015_Pu_Wu}. It was also pointed out that such instabilities preferentially depleted more pairs narrow of the MMR \citep{2015_Pu_Wu, 2019Dong-Hong_Wu}. Compact resonant chains may also get destroyed due to dynamical instabilities \citep[e.g.,][]{2017Izidoro, 2021_Izidoro,2020_Matsumoto,2020_Pichierri,2022_Goldberg}.

In this paper, we revisit the problem and further explore the idea proposed by CF15. CF15 argued that while a significant gas disk is present, disk-driven migration can trap planet pairs into MMRs. After the gas disk is sufficiently depleted, these trapped planet pairs dynamically interact with nearby planetesimals from a residual planetesimal disk. CF15 specifically considered planet pairs initially trapped into the $2:1$ MMR. They showed for a wide range of disk profiles that these interactions can break the $2:1$ MMR and if so, the planet pairs diverge creating large offsets. They argued that this process can be responsible for creating the asymmetry in the distribution of near-resonant planet pairs. We improve and expand the scope of this study in several key aspects. Since CF15 only considered planet pairs initially trapped in $2:1$ MMR, their results do not explore the effects of planetesimal scattering on planet pairs that are initially narrow of the nominal MMR position. As a result, CF15's results are unable to explain the deficit of systems narrow of period ratio $2:1$. Moreover, CF15 did not explore the effects of planetesimal scattering near $3:2$ at all. In our study, we consider planet pairs near $2:1$ as well as $3:2$; we also consider planet pairs that are initially trapped in MMRs as well as those that are not. While CF15 considered specific mass ratios between planet pairs in a grid, we consider masses that are directly motivated by the observed planet pairs in the exoplanet database. Furthermore, while CF15 placed the planet pairs in a way such that the inner planet is near $0.5\,\au$, we assign orbital separations guided directly by the observed exoplanet pairs. In order to limit the computational cost, we restrict ourselves only to study a single planetesimal disk profile which initially follows the profile of the minimum-mass solar nebula, since CF15 showed that the results of planet-planetesimal scattering do not qualitatively depend on the adopted disk profile. Note that the main ingredient in CF15 as well as this study is that during planet formation a significant fraction of solids may not grow into planets. As a result, after gas dissipates, (almost) fully formed planets coexist with the debris of planet formation \citep[e.g.,][]{2012_Schlichting, 2020_Mulders}. Also note that in the context of planet formation, solids can have a wide variety of names, such as pebbles, debris, rocks, planetesimals, and protoplanets, depending on their sizes, but the boundaries between them are blurry. For simplicity, throughout the paper, we will use the word `planetesimals' rather loosely and mean all left over solids that did not grow into planets or protoplanets. We will provide a more physically motivated definition of the upper limit of mass for each of the so-called planetesimals later on.

We present our work as follows. In \autoref{sec:num} we describe the physical picture we are trying to simulate, the initial properties of the planets and planetesimal disks, and the details of our numerical models. In \autoref{sec:res} we present our key results. We summarise and conclude in \autoref{sec:discuss}.

\section{Numerical Setup} \label{sec:num}
The core-accretion paradigm of planet formation suggests that the newly born planets are embedded in a protoplanetary disk. The disk is initially made of mostly gas and dust. While significant uncertainties remain in how planet cores grow from dust, it is generally assumed that cores grow in the disk and some of these cores can also accrete significant gas to become giant planets \citep[e.g.,][]{2004_Ida_Lin, 2010_Armitage_book, 2018_Mordasini, 2019_Lee}. It is also generally expected that by the time the gas disk dissipates, some of the solids go into the formation of planets or planetary cores, while the rest of the solids remain as left-over of the planet formation process. This debris of the planet formation process later on either gets accreted by the planets or the star, get cleared away by dynamical scattering with the planets \citep[e.g.,][]{2006_Higuchi_Kokubo_Mukai}, or erode through collisional grinding \citep[see reviews by][]{2008_Wyatt, 2010_Krivov, 2018_Hughes}. For simplicity, as mentioned before, we will call all solids that do not grow into planets (or planetary embryos, or protoplanets depending on the definition) at the time of gas disk dispersal as planetesimals.

While the gas disk is sufficiently massive, in particular, while the timescale for dynamical instability is long compared to the damping timescale from the gas disk \citep[e.g.,][]{Matsumura2010}, dynamical excitations cannot grow and the possibility of dynamical interactions between planets and planetesimals is reduced. The torques raised by the gas-disk, depending on the migration speed and resonance strength may also trap a fraction of planet-pairs into MMRs  \citep{2002LeePeale,Papaloizou2005, Batygin_2015, Deck2015, Delisle_2015}. During this damped dynamics phase, some planet-planetesimal interactions still may occur if the instability timescale is shorter compared to the ever-increasing damping timescale. Eventually, the gas-disk depletes completely, leaving behind the planets and a disk of residual solids. Beginning from when the gas-disk is sufficiently depleted such that the damping timescale is long compared to the timescale of subsequent dynamical scatterings, the planets and planetesimals can freely interact and dynamical excitation of orbits can freely grow. 

While capturing this general scenario fully is not possible with present-day computational abilities, it is quite likely that at the time of gas dispersal, fully formed or almost fully formed planets coexist with a sea of planetesimals, which are conceivably the progenitors of what we see, after billions of years of evolution, as the asteroid belt or the Kuiper belt in our solar system \citep[e.g.,][and references therein]{1996_Stern, 2007_O'Brien}. The density distribution of planetesimals near the planets is hard to constrain and likely depends on the planet formation process. Since there are little constraints on the details of this stage of the evolution, we make several simplifying assumptions while keeping the physical process we have in mind unaltered to make this problem tractable.

We notice that the interactions that shape the planetary orbits can be easily divided into two clearly separate regimes with some possible interim phase; while the gas-disk is sufficiently massive, gas-disk--planet interactions are expected to dominate in shaping the orbital properties of the planets; on the other hand, when the gas disk is sufficiently depleted, the planet-planetesimal scattering should dominate the evolution of planetary orbits. There may be some interim phase where some planet-planetesimal scattering is allowed for which the instability timescale is too short compared to the depleting gas disk. Note that, after gas disk depletion, planet-planet scattering is also a possibility \citep[e.g.,][]{Rasio_Ford_1996,Lin_1997, Ford_2007, levison2007, Chatterjee_2008}. However, planet-planet scattering is expected to perturb orbits too strongly and is not expected to give rise to the subtle observed abundance asymmetry in the period-ratio distribution across the MMRs \citep[e.g.,][]{Chatterjee_2008, Chatterjee&Ford2015}.

Each of our simulated systems consists of two planets orbiting their host star embedded in a dynamically consistent (see \autoref{sec:num2}) planetesimal disk. The same calculations can be repeated with systems with more than two planets too. However, a higher number of planets in the system simply makes the dynamics more complex making it harder to interpret the results. Nevertheless, our calculations should be applicable to planetary systems with higher planetary multiplicities if it can be assumed that the system is pair-wise dynamically stable. We focus on how the planet-planetesimal disk interactions alter the period ratios of planet pairs when they are either trapped in an MMR or in the vicinity of an MMR. Throughout our study, we use the hybrid integrator module \mercurius\ available in the \rebound\ simulation package \citep{ReinREB2012, MERCURIUS2019}. 

\subsection{Creation of Initial Conditions}
\label{sec:ICs}
In this section we describe the details of how we create and approximate initial conditions expected to be present at the time of gas-disk dispersal. Because of the complex nature of the condition, we create the initial conditions in several steps. These steps are described below.

\subsubsection{Stage 1 : Planet and star properties, initial orbits} 
\label{sec:num1}
By design our setup is agnostic to the formation process. Thus, instead of trying to build the planets ground up assuming any specific formation process, we start with two planets that are almost fully formed with minor opportunities for further growth from accretion of planetesimals. In order to preserve the orbital architectures and various scalings present in real systems, we draw the planet pairs and the host stars for our models directly from the observed properties of adjacent planet pairs in multiplanet systems. The observed exoplanet data used are taken from NASA's Exoplanet Archive \citep[][]{2013Akeson_NEA}.\footnote{\url{http://exoplanetarchive.ipac.caltech.edu}, \citep{2013Akeson_NEA} updated on August 12, 2022.} Without applying any filters, out of the $837$ multiplanet systems we can find $1272$ adjacent planet pairs in total. We exclude systems around giant stars with surface gravity $\log(g/\unitacc)\leq3$, which reduces the sample size to $820$ multiplanet systems with $1254$ adjacent planet pairs. We further exclude systems with missing information on critical properties, such as both mass and radius of the planets are unknown, or host star properties are unknown. In addition, we restrict our study to planet-pairs with sub-Neptunes ($R_{\rm{p}} < 4\,R_{\Earth}$). While larger and more massive planets may also exist in real systems, changing their orbits requires proportionally higher mass in interacting planetesimals (e.g., CF15). These cuts reduce the number of planet pairs to $905$ distributed in a broad range of period ratios. Since the focus of this study is to explain the observed dearth (excess) of systems narrow (wide) of period ratios $3:2$ and $2:1$, we further select only the planet pairs with period ratios below $2.5$, which leaves us with $582$ pairs in $377$ systems. In this chosen sample, $175$, $180$, and $219$ pairs are in systems with 2, 3, or more than 3 known planets, respectively. The asymmetry across period ratios $3:2$ and $2:1$ for pairs in two-planet systems and systems with higher multiplicities do not show statistically significant differences. Hence, although we simulate two-planet systems only, we keep all observed adjacent pairs to improve the statistics in observed systems.

In a large fraction of our selected sample, precise mass measurements are not available, we follow the following scheme based on the reported planet mass ($m$) and the corresponding error ($\delta m$)
\begin{itemize}
  \item $\delta m/m\leq0.3$: We use the nominal mass value reported in the database directly. 
  \item $\delta m/m>0.3$ or either mass or radius measurements are available (but not both): We use the M-R relationship from \citet{2017Chen_Kipping} to estimate the mass (radius) based on the reported radius (mass) in the database. 
  \item Average planet density $\rho_{\rm{p}} > 60\,\gcc$, which corresponds roughly to the density of a $100\,M_{\Earth}$ iron planet \citep{Fortney_2007}: We use the M-R relationship \citep{2017Chen_Kipping} to estimate mass.
\end{itemize}

The last condition is enforced, because in some observed systems the reported masses are from dynamical constraints. They are generally overestimated and may create unphysical models. We further make sure that all planet masses are lower than the mass of Neptune. At last, we have all necessary physical properties including the mass and radius of $574$ planet pairs around $374$ host stars.

We now proceed with initializing their orbital properties. Note that, we cannot directly import the observed orbital periods ($\porb$) for our planets since these would potentially change due to planet-planetesimal scattering. On the other hand, from CF15 we know that the amount of migration via planet-planetesimal scattering is limited. Hence, to ensure that the planet pairs are roughly at $\porb$ similar to the observed, we directly import the observed $\porb$ of the outer planet ($P_2$). We divide our models into two major subsets and assign the rest of the orbital properties in the following way.

\begin{itemize}
    \item Initially non-resonant pairs: In this subset, all planet pairs are initially non-resonant but are near period ratios $3:2$ or $2:1$. We assign $P_1$ such that $\epsilon$ is distributed uniformly between $-0.14 \le \epsilon \le +0.12$. The eccentricities (inclinations) of the planetary orbits are drawn according to the Rayleigh distribution with mean 0.04 \citep[0.024;][]{Xie201604692}. Orbital phase angles are chosen uniformly in their full ranges. Once the orbits are created, we ascertain that the resulting orbits are stable against planet-planet scattering and orbit crossing using the generalized stability criteria \citep{1993Gladman, 2018Petit}. If we find that a planet pair is unstable we re-draw the orbital properties until we find a stable orbital configuration. This configuration is further integrated for $100\ P_2$ to ensure orbital stability. We collect all relevant properties for these planet pairs at the end of these integrations.

    \item{Initially resonant pairs:} In this subset we create initially resonant planet pairs in the following way. We first assign $P_1$ such that the planet-pairs lie just outside the nominal position of the MMR, e.g., at $P_{2}/P_{1} = 1.52$ for the $3:2$ MMR and at $P_{2}/P_{1} = 2.05$ for the $2:1$ MMR. Initially, both planets' orbits are circular and co-planar. 

    We apply a slow inward migration on the outer planet with migration timescale, $T_{a} = 10^{6} P_{2}$ and eccentricity damping timescale, $T_{e} = 10^{4} P_{2}$ for a duration of $T_{\rm{mig}} = 2 \times 10^{4} P_{2}$ to trap the planets in MMR. We ensure trapping into MMR via libration of the resonant angles for $(p+1)/p$ MMR-
    \begin{eqnarray}
        \theta_{1} & = & {(p+1)\lambda_{\rm{o}} - p\lambda_{\rm{i}} - \varpi_{\rm{i}}} \nonumber\\
        \theta_{2} & = & {(p+1)\lambda_{\rm{o}} - p\lambda_{\rm{i}} - \varpi_{\rm{o}}} \nonumber\\
        \theta_{3} & = & {\varpi_{\rm{i}} - \varpi_{\rm{o}}},
    \end{eqnarray}
    where $\lambda$ ($\varpi$) is the mean longitude (longitude of pericenter) and the subscript $\rm{i}$ ($\rm{o}$) refers to the inner (outer) planet. If the adopted timescales, $T_a$, $T_e$, and $T_{\rm{mig}}$, results in overshooting the resonance instead of trapping, we adjust $T_{\rm{mig}}$ to ensure trapping (needed in $43$ systems out of $2600$ across both the MMRs). Note that, the exact choice of these timescales is not directly relevant for this study. We are simply interested in creating a large set of initially resonant planet pairs with properties that are observationally motivated. We collect all relevant properties for these planet pairs at the end of the resonance trapping integrations.
\end{itemize}

\subsubsection{Stage 2 : Introduction of planetesimal disk} \label{sec:num2}

We adopt an initial planetesimal disk where the surface density $\Sigma(r) \propto r^{-3/2}$, similar to the profile of the minimum-mass solar nebula \citep[MMSN, e.g.,][]{1977Weidenschilling, 1981Hayashi}, where $r$ denotes the distance from the star. Note that we do not claim that the density profile of planetesimals are always expected to be continuous or given by a power-law. In fact, it is expected that the solid delivery may be significantly dependent on the planet formation scenario. For example, in case of significant migration, the planetesimal profile may not be continuous and may instead have overdensities at resonance locations \citep{2003_Wyatt}. On the other hand, in-situ formation models envision that there may have been a wide diversity of profiles for solids both in the exponent as well as the normalisation \citep[e.g.,][]{2012_Hansen_Murray, 2013_Hansen_Murray, 2013_Chiang_Laughlin, 2016_Moriarty, 2020_MacDonald}. In contrast, the inside-out formation model predicts that there is a continuous supply of solids streaming in from the outer disk \citep{2014_Chatterjee_Tan, 2015_Chatterjee_Tan, 2016_Hu}. CF15 studied the effects of different disk profiles and showed that a change in the disk profile simply changes the fraction of planetesimals close enough to interact with the planets and does not change the qualitative outcome of these interactions. Hence, to keep this already complex problem numerically tractable, we start with the MMSN-like disk profile and do not explore other disk profiles. Nevertheless, it can potentially be interesting to investigate different initial disk profiles in a future work.

To avoid any unpredictable edge effects we set the inner (outer) edge of the disk at $P_1/3$ ($3P_2$), sufficiently far away from the positions of the planet pair. The orbital elements of the planetesimals (e.g. eccentricity, inclination, and other phase angles) are assigned in the same way as the non-resonant planet pairs (\autoref{sec:num1}). We treat the planetesimals as pseudo test-particles to limit computational cost; while planetesimal-planet interactions are taken into account, planetesimal-planetesimal interactions are ignored. This is an approximation often adopted by similar studies of dynamical systems \citep[][]{2005_Tsiganis_NiceModel, 2012_Raymond, 2014_Bonsor, 2014_Izidoro, Chatterjee&Ford2015, 2020_Mulders} and ignores the possibility of collisional growth and fragmentation of the planetesimals themselves. Since we are interested in the effects of a sea of small solids on the planets, this approximation is acceptable for our study. Furthermore, we expect that the left-over solids will not quickly grow themselves in the presence of the planets, rather they will either get scattered or accreted by the planets.

The size of the planetesimals are found using density $\rhopl=4\,\gcc$ and equal mass given by $\mpl = \mdisk/\Npl$, where $\mdisk$ ($\Npl$) denotes the total initial mass (number) of planetesimals in the disk. Although, in reality, the solid disk may have a variety of sizes \citep[e.g.,][]{2009_Morbidelli, 2011_Weidenschilling, 2013_Schlichting}, we do not expect this simplification changes the nature of the physical process we are interested in. We parameterise the total disk mass to be a function of the total planet mass $\mplanet = m_1+m_2$; $\mdisk = k \times \mplanet$, where $k$ is a constant. We explore $k = 0.1$, $0.5$, $1$ and $2$. It is important to note, that the total disk mass $\mdisk$ is actually not an interesting parameter in this problem and only relevant for our bookkeeping. What really matters is the total mass of planetesimals that are close enough to the planets to dynamically interact with them (see discussion in \autoref{app:md_vs_mint}).

The choice of $\Npl$ in our simulations is somewhat arbitrary. We intend to investigate how the orbital properties of the planets evolve as a result of the cumulative effects of a large number of weak interactions from a sea of low-mass objects. Thus, on one hand, the higher the $\Npl$ the better. On the other hand, the computational cost scales as $\Npl$ forcing us to make a pragmatic choice. We choose $\Npl$ by demanding that the fractional change in the semi-major axis caused by a single planet-planetesimal interaction should always be small, $\Delta a/a \leq 10^{-3}$. Since $|\Delta a / a|_{\rm{max}} \sim \mpl / m$ for a single planet-planetesimal encounter, this requirement ensures that the mass of a planetesimal never exceeds $1/1000$th of the mass of the lower mass planet in the simulated system. As a fiducial value, we choose $\Npl=5000$. In a handful of systems with disparate planet masses, our fiducial $\Npl=5000$ does not satisfy the condition that $\mpl<10^{-3}$ times the lower-mass planet. In these cases we increase $\Npl$ as required, up to $\Npl=10^4$.  Our tests indicate that as long as $\mpl$ is sufficiently small compared to $\mplanet$, the results we are interested in do not change and we do capture the cumulative effects of a large number of weak encounters as expected from left-over solids interacting with the planets (see more detailed discussion in \autoref{app:mpl_effect}). This consideration indirectly imposes a constraint on the maximum planet-planet mass ratio we can explore for a given $\mdisk/\mplanet$. For example, for $\mdisk/\mplanet =1.0$, the maximum mass ratio between the planet pairs is $\approx 9:1$, hence, we restrict the maximum mass ratio to this value. Using this additional constraint, we are left with $548$ planet pairs for $\mdisk/\mplanet =0.1, 0.5, 1.0$ and $529$ pairs for $\mdisk/\mplanet=2.0$ models. Discarding extreme mass ratio systems should not affect our overall results significantly since this is a small fraction of all systems. It is also expected that adjacent planet pairs are more similar to each other than two randomly drawn planets \citep[e.g.,][]{Weiss_2018a}; we suspect that many of the apparently high mass ratios may not be real and is a result of using the nominal M-R relationship which, in reality, has large spreads \citep[e.g.,][]{Chatterjee&Ford2015,Wolfgang2016,2017Chen_Kipping}.

\begin{figure}[htb]
\plotone{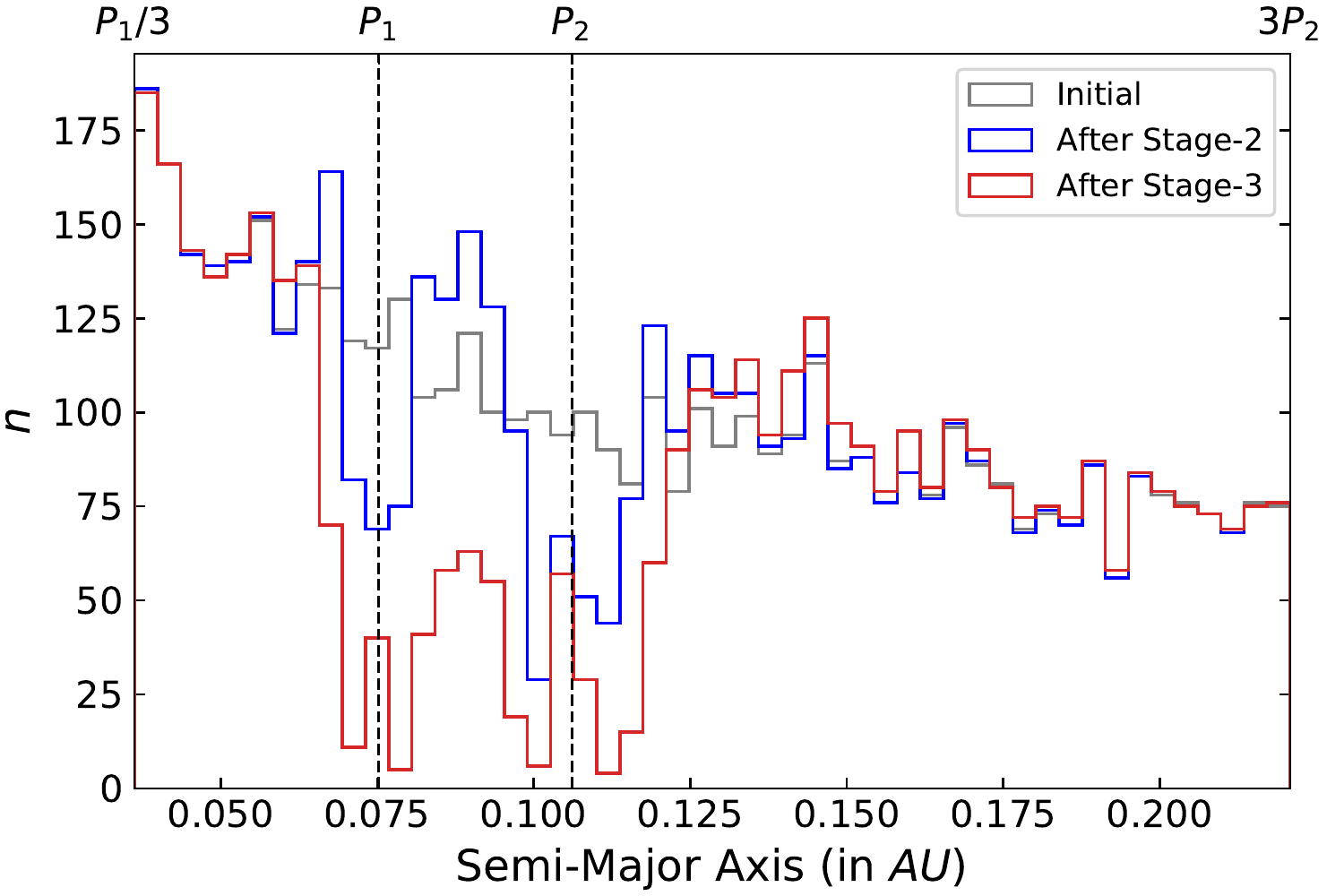}
\caption{Evolution of the planetesimal disk profile as an example. The initial profile (grey), the profile after Stage-2 (blue), and the final profile (red) are shown. The positions of the planets after Stage-2 are marked by the vertical dashed lines. This example system consists of a non-resonant planet pair ($m_1/M_{\Earth}=2.1$, $m_2/M_{\Earth}=5.48$) around a $0.84 M_{\sun}$ star, embedded in a planetesimal disk with initial $\mdisk/\mplanet = 0.5$.
\label{fig:planetesimal_dist_stages}}
\end{figure}

In a real system emerging out of a gas disk, the planetesimal disk profile is not expected to remain a pure power-law. Even in the presence of damping from the gas disk, significant dynamical clearing is expected for planetesimals that are in orbits unstable on short timescales. To model this, we introduce the planetesimals as zero-mass particles and gradually increase their mass to the desired value over a timescale $\tramp=10^3P_2$. Thus, at the beginning, the planets perturb the orbits of the nearby planetesimals without any effect on their own orbits. As a result, only planetesimal orbits that are dynamically consistent with the orbits of the planets remain. The gradual increase of $\mpl$ also ensures that the systems do not get shocked by a sudden introduction of a massive planetesimal disk. The choice of $\tramp$ is somewhat arbitrary but it is easy to understand how $\tramp$ affects the evolution. A small $\tramp$ may shock the system and potentially create instabilities early. A large $\tramp$ can lead to a lot of planet-planetesimal interactions before the planetesimals reach the full predetermined mass. In our tests, we find that as long as $\tramp\gg P_{2} $, the choice of $\tramp$ does not alter the planets' orbits significantly and the exact value has little effect on the final results. At the end of this clearing stage in each system, we obtain two planets embedded in a {\em dynamically consistent} disk of planetesimals. \autoref{fig:planetesimal_dist_stages} shows an example of a planetesimal disk and the embedded planet pairs at the time of the introduction of the planetesimal disk (grey) and when the planetesimals have attained their full mass (blue). The planet positions are given by the vertical lines. At the end of stage 2, we are left with two planets embedded in a disk of planetesimals with structures in the density profile that is dynamically consistent with the presence of the planets. We envision that as long as the planet formation process is wasteful in use of the solid reservoir, planetary systems emerging out of a gas disk may have conditions that are dynamically similar to our setup at the end of stage 2.

We create 1300 systems of planet-pairs embedded in a dynamically consistent planetesimal disk, each around initial period ratios of $3:2$ and $2:1$ for the initially non-resonant set. For the set of initially resonant planet-pairs, we again create $1300$ systems each for the $3:2$ and the $2:1$ MMRs. Thus, in total, we have $5200$ planet pairs initialised either in initially resonant orbits or in initially non-resonant orbits. This is done for each $\mdisk/\mplanet$, which means that we have simulated a total of $5200 \times 4$ systems.

\subsection{Stage 3 : Evolution of Embedded Planet Pairs} \label{sec:num3}

We simulate the planet pairs embedded in a dynamically consistent disk for $10^5P_2$ to study how the planet pairs evolve due to the cumulative effect of many stochastic small interactions with the planetesimals. The frequency of interactions between the planets and planetesimals is high initially and gradually decreases with time, as expected. The majority of these interactions happen earlier than our chosen stopping time. Moreover, we verify that longer integrations do not alter the results in a statistically significant way.

\section{Results} \label{sec:res}
In this section, we first describe the typical evolution of initially resonant and non-resonant planet-pairs under the influence of a planetesimal disk using example systems. Afterwards, we describe how these interactions shape the final $\epsilon$-distributions of the whole ensemble of planet pairs we simulate.

\subsection{Typical evolution of planet pairs} \label{sec:resnr}
In a typical system, the planetesimals interact with the planets stochastically. Similar to the findings of CF15, a wide range of outcomes of individual planet-planetesimal interactions are possible. No single interaction produces any significant perturbations to the planetary orbits. However, the cumulative effect of a large number of encounters typically increases the ratio of the planetary orbital periods.\footnote{For initially non-resonant planets, $\epsilon$ increases in $\sim 97\%$ ($\sim 65\%$)of our simulated systems near $3:2$ ($2:1$). The corresponding fraction for systems initially in the $3:2$ ($2:1$) MMR is $\sim 96\%$ ($\sim 92\%$)}.

\subsubsection{Initially non-resonant pairs} \label{sec:typicalnonres}

\begin{figure}[htb]
\plotone{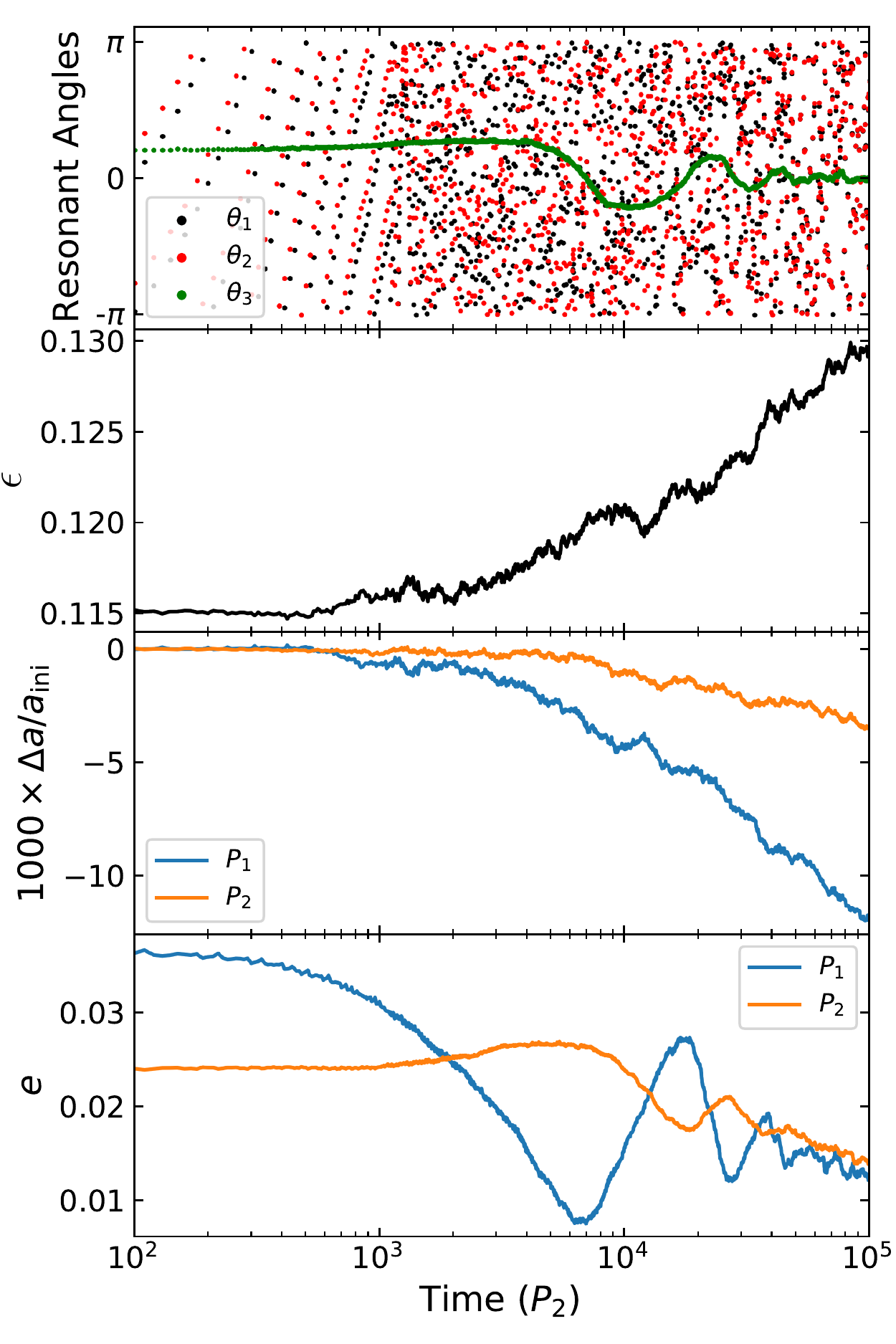}
\caption{Orbital evolution of a typical non-resonant planet pair (same system as in \autoref{fig:planetesimal_dist_stages}) initially wide of $3:2$ ($\epsilonini=0.115$), embedded in a planetesimal disk of initial $\mdisk/\mplanet = 0.5$. From top to bottom, the panels show the evolution of the resonant angles corresponding to the $3:2$ MMR, the offset from the nominal resonance position $\epsilon$, the change in semi-major axes normalised by the initial values, and the orbital eccentricities, respectively. Time is in units of the outer planet's orbital period, $P_2$. Due to planetesimal interactions $\epsilon$ increases. The resonant angles circulate throughout the simulation. In this example, the semi-major axes of both planets decrease. The eccentricities decrease over time due to planetesimal interactions.
\label{fig:evo_nr}}
\end{figure}

\autoref{fig:evo_nr} shows the evolution of the orbital properties for an initially non-resonant planet-pair under the influence of planet-planetesimal interactions as an example. Note that throughout the paper, $t=0$ indicates the start of stage\ 1 in our initial setup. As a result, the effects of the planet-planetesimal scattering are fully underway starting from $1.1\times10^3\ P_2$. In this example system, the planets initially are wide of the nominal $3:2$ MMR with an initial offset, $\epsilonini=0.115$. 
Throughout the evolution, the resonant angles corresponding to $3:2$ commensurability circulates (top panel). While at the beginning the period ratio remains roughly unchanged, it monotonically increases after $t\sim10^3\ P_2$ and reaches $\epsilon=0.129$ at the integration stopping time (second panel from top). In this example, both planets migrate inwards, however, the inner planet migrates more than the outer (third panel from top). In our simulations, we find that several combinations are possible and the typical migration direction leading to an increase in $\epsilon$ varies depending on the initial resonance status and period ratio. For example, in the case of initially non-resonant planets near $3:2$, the inner planet typically migrates inwards while the outer planet migrates outwards. In contrast, near $2:1$, usually, both planets tend to migrate inwards, while the inner planet migrates more than the outer. The orbital eccentricities for both planets are damped due to planet-planetesimal scattering (bottom panel). At the integration stopping time the planets open large enough cavities in the planetesimal disk (\autoref{fig:planetesimal_dist_stages}) and planet-planetesimal scattering becomes infrequent. All planet pairs initially wide or narrow of $3:2$ or $2:1$ period ratios show very similar evolution if the increase in period ratios does not push them very close to the resonance.

\begin{figure}[htb]
    \plotone{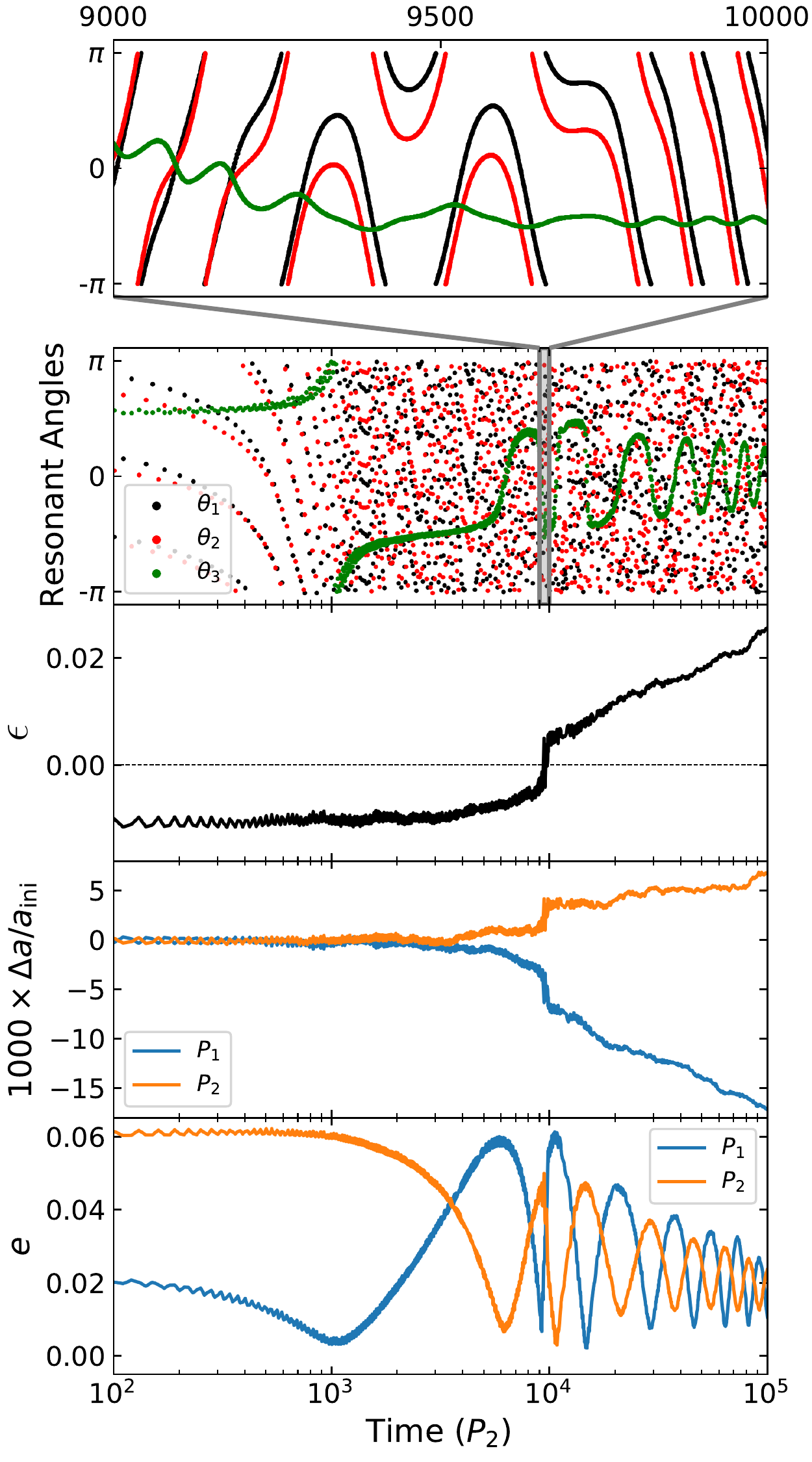}
    \caption{Same as \autoref{fig:evo_nr} but for planet pairs ($m_1/M_{\Earth}=2.27$, $m_2/M_{\Earth}=3.02, M_{*}/M_{\sun} = 0.78$) initially narrow of $3:2$ ($\epsilonini=-0.01$) embedded in an initial planetesimal disk of $\mdisk/\mplanet = 0.5$. The offset increases rapidly as the planets approach $\epsilon=0$, the planets overshoot the $3:2$ MMR and get deposited wide of $3:2$. Afterwards, $\epsilon$ increases monotonically. The resonant angles circulate throughout the simulation except during the short time taken by the planets to overshoot the $3:2$ resonance, shown in the zoom-in panel at the top. In this particular example, the inner planet migrates inwards and the outer planet migrates outwards. Both eccentricities are damped.
    }
    \label{fig:eps_evo_nr_cross}
\end{figure}

We find a qualitatively different evolution if $\epsilonini<0$ and the planet pairs cross the nearest first-order MMR ($3:2$ or $2:1$) as the period ratio increases. \autoref{fig:eps_evo_nr_cross} shows the evolution of two planets with initial period ratios narrow of the $3:2$ MMR as an example. As the planets reach sufficiently close to the nominal period ratio for the $3:2$ resonance, $\epsilon$ suddenly jumps across zero, the nominal position of the resonance. As a result, the planets overshoot the $3:2$ resonance and get deposited significantly wide ($\epsilon\gtrsim 0.005$) of $3:2$. The resonant angles circulate almost throughout the evolution except for a short time during the resonance crossing when they librate. This jump in $\epsilon$ is likely a result of resonant repulsion \citep{Lithwick_2012}. After this jump, $\epsilon$ continues to monotonically increase as long as the planets have access to planetesimals to interact with. In general, systems that are close enough to an MMR on the narrow side, in presence of a sufficiently massive planetesimal disk jump across the resonance and this jump is primarily responsible for the observed paucity of systems narrow of the $3:2$ and $2:1$ MMRs. The evolution of systems narrow of the $2:1$ MMR is very similar with some differences showing up at the population level depending on the relative strengths of the $3:2$ and $2:1$ MMRs. More discussion on this later.

\subsubsection{Initially resonant pairs}
\label{sec:typicalres}
\begin{figure}[htbp]
\plotone{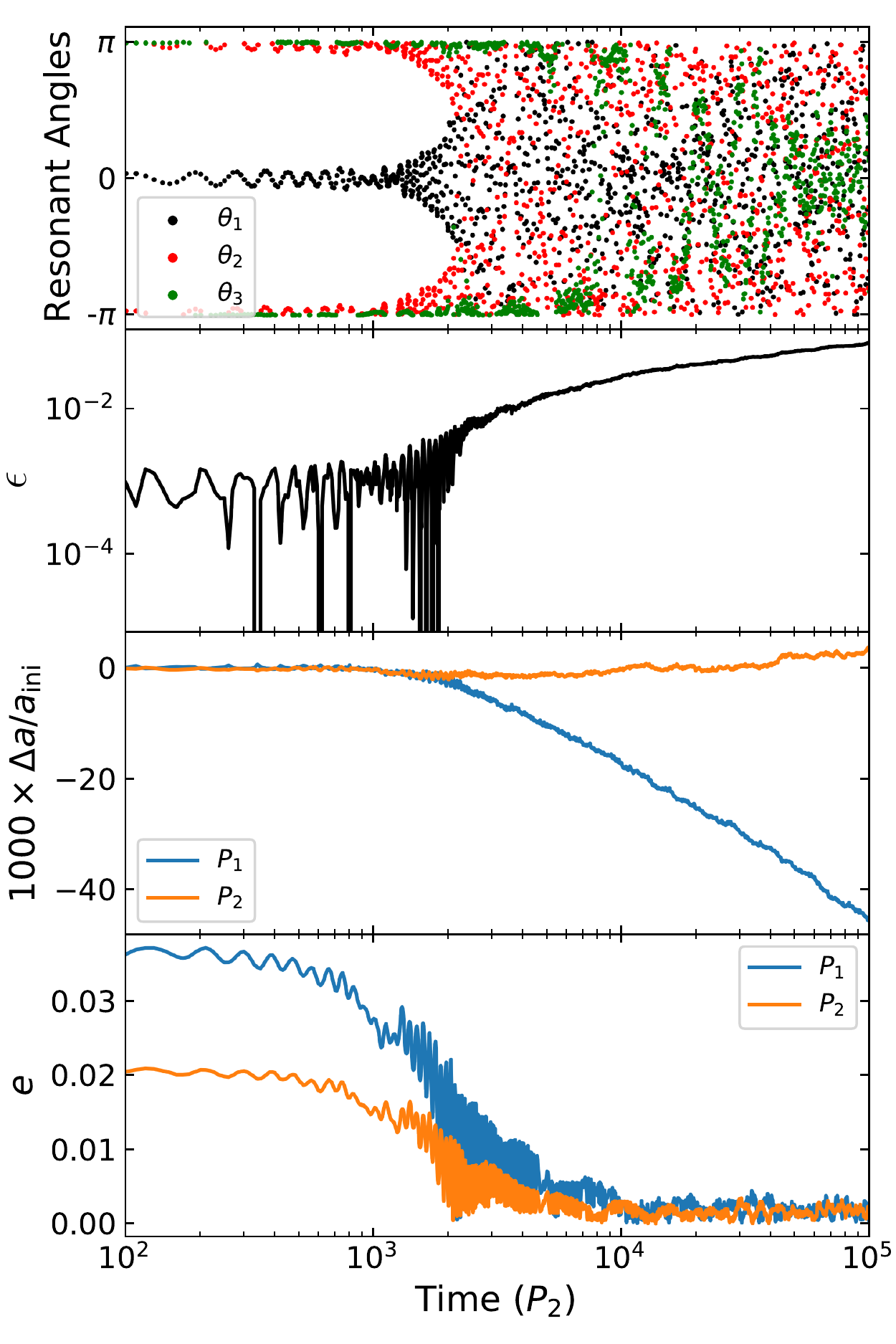}
\caption{Same as \autoref{fig:evo_nr} but for two planets ($m_1/M_{\Earth}=8.6$, $m_2/M_{\Earth}=16.6, M_{*}/M_{\sun} = 1.04$) initially in the $3:2$ MMR embedded in a disk of initial $\mdisk/\mplanet = 0.5$. The resonant angles librate until $t\sim3\times10^3\,P_2$. Afterwards, planetesimal interactions break the resonance and the resonant angles start circulating. While in resonance, planetesimal interactions lead to stochastic changes and $\epsilon$ fluctuates while remaining small ($\epsilon\lesssim10^{-2}$). After resonance breaks, $\epsilon$ increases monotonically to reach $\epsilon\sim0.1$. In this system, the inner planet migrates inward while the outer planet migrates outwards. All eccentricities are damped. }
\label{fig:evo_32_r}
\end{figure}

The evolution of planet pairs initially in $2:1$ MMR and embedded in a planetesimal disk was discussed in CF15 in detail. In this study, we consider planet pairs initially trapped in $2:1$ as well as $3:2$ MMR. The behavior of planet pairs initially in resonance under the influence of a planetesimal disk is similar to non-resonant planet pairs except that the interactions first need to break the resonance before affecting the essentially monotonic increase in $\epsilon$. \autoref{fig:evo_32_r} shows the evolution of orbital properties for a planet pair initially trapped in the $3:2$ MMR embedded in a planetesimal disk with an initial $\mdisk/\mplanet = 0.5$ as an example. The resonant angles librate until $t\sim2\times10^3\ P_2$ (top panel) indicating that the planet pair is initially trapped in the $3:2$ MMR. While the planets are trapped in the resonance, the evolution of $\epsilon$ is random and fluctuates with a small ($\lesssim 10^{-3}$) amplitude. Once the resonance is broken at $t\sim2\times10^3\ P_2$, easily identified by the circulation of the resonant angles, the planet pair practically becomes a non-resonant planet-pair wide of the MMRs as discussed in \autoref{sec:resnr} (\autoref{fig:evo_nr}). Then the planets undergo divergent migration as long as enough planetesimals are available in the vicinity to interact. In this particular example, the two planets migrate inward together while they are trapped in the resonance. However, once the resonance breaks, the inner planet continues the inward migration while the outer planet migrates outward. In general, for the case of initially resonant planet pairs similar to the case of initially non-resonant pairs, $\epsilon$ increases because of either both planets migrating inwards (inner planet migrates more than the outer) or the inner planet migrating inwards and the outer planet migrating outwards.

In the example of \autoref{fig:evo_32_r}, the eccentricities of both planets decrease to very low values due to planet-planetesimal interactions, similar to the initially non-resonant planet pairs (bottom panels in \autoref{fig:evo_nr}, \autoref{fig:eps_evo_nr_cross} \& \autoref{fig:evo_32_r}). In general, a higher total mass in interacted planetesimals ($\minteract$) leads to more damping. For example, while the mean eccentricity for initially non-resonant planets is $\bar{e}=0.04$ at the beginning of Stage\ 1 (\autoref{sec:num1}), it reduces to $\bar{e}=0.034$, $0.018$, $0.009$, and $0.004$ in the final snapshot for our models with $\mdisk/\mplanet=0.1$, $0.05$, $1.0$, and $2.0$, respectively. Similarly, in case of initially-resonant planets, $\bar{e}$ reduces from $\approx0.02$ at the end of Stage\ 1 to $\approx0.01$ and $<0.004$ in models with initial $\mdisk/\mplanet=0.1$, $\geq0.5$, respectively. Similar to eccentricities, the mutual inclinations ($\imut$) also decrease. For example, initially, $\bar{\imut}=0.024$ for non-resonant systems. The final $\bar{\imut}=0.014$ ($\lesssim0.003$) for models with $\mdisk/\mplanet = 0.1$ ($\ge0.5)$. The resonant systems in our models were coplanar at the end of Stage\ 1. The interactions break the coplanarity, but $\imut$ remains very low, $\bar{\imut}\lesssim0.001$. Overall, planetesimal interactions make the two-planet system more stable by reducing $e$ and increasing period ratios.

\subsection{Distribution of \texorpdfstring{$\epsilon$ }{Lg}}
\label{sec:epsilon}

\begin{figure}[htb]
\plotone{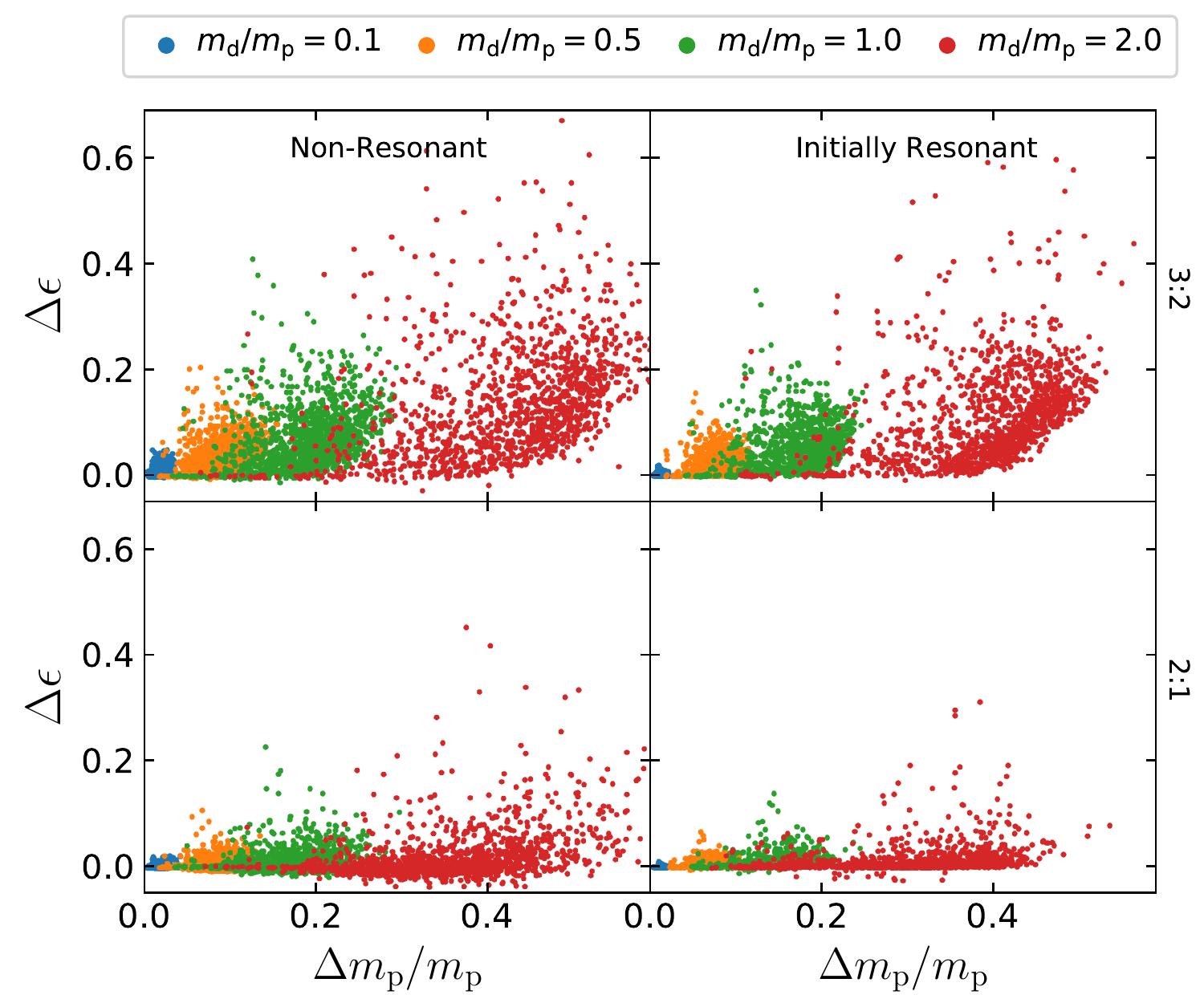}
\caption{The change in offset, $\Delta \epsilon$ vs the fractional gain in total planet mass ($\fracdelm$). Planet-planetesimal collisions lead to increase in $\mplanet$ which we use as a proxy for strong planet-planetesimal interactions. The top and bottom panels show systems near $3:2$ and $2:1$, respectively. The right and left panels show systems that are initially resonant and not in resonance, respectively. Different colors denote systems modeled with different initial $\mdisk/\mplanet$ (see legend).  We find that $\Delta \epsilon$ is correlated with $\fracdelm$. Furthermore, as expected, the higher the disk mass, the higher the $\fracdelm$ and $\Delta \epsilon$.
\label{fig:del_eps_comb}}
\end{figure}

The change in offset, $\Delta\epsilon\equiv\epsilon_{\rm{fin}} - \epsilon_{\rm{ini}}$, where, $\epsilon_{\rm{fin}}$ ($\epsilon_{\rm{ini}}$) denotes the final (initial) $\epsilon$, depends on the number of strong planet-planetesimal interactions in a particular system. As mentioned earlier, the overall $\mdisk$ is not interesting for our problem, instead, $\minteract$ is the key quantity. For the example system shown in \autoref{fig:planetesimal_dist_stages}, $\minteract/\mplanet \approx 0.09$, whereas, $\mdisk/\mplanet=0.5$. Indeed, we find that for various different $\mdisk/\mplanet$, if $\minteract/\mplanet$ is roughly similar, the $\epsilon$ evolution too remains unchanged (see \autoref{app:md_vs_mint} for more details).

It is impractical to track every strong scattering between the planets and planetesimals for all our simulations because of the unreasonably large data size. Instead, we use the growth of the total planet mass ($\Delta\mplanet$) via planet-planetesimal collisions as a proxy for the number of strong planet-planetesimal encounters. In \autoref{fig:del_eps_comb}, we show $\Delta\epsilon$ as a function of $\Delta\mplanet/\mplanet$ for all planet pairs in our simulations. Different colors denote models using different initial $\mdisk/\mplanet$. The top and bottom panels show systems near the $3:2$ and $2:1$. The left and right panels show models with initially non-resonant and resonant systems. Several trends become apparent. The higher the level of interactions (and as a result, growth of planet mass via planet-planetesimal collisions), the larger the $\Delta\epsilon$. Moreover, for any particular $\Delta\mplanet/\mplanet$ there is a large spread in $\Delta\epsilon$, which illustrates the stochastic nature of the evolution. Clearly, it is not possible to draw one-to-one correspondence between $\Delta \epsilon$ in a particular system and the total amount of interactions.\footnote{For a longer discussion on this aspect see CF15.} There is significant overlap between systems with different initial $\mdisk/\mplanet$. Moreover, we find that for the same $\mdisk/\mplanet$, systems near $P_2/P_1=3:2$ exhibits a statistically higher $\Delta\epsilon$ compared to those near $P_2/P_1=2:1$. Near both $2:1$ and $3:2$, the initially resonant planet pairs show a smaller $\Delta\epsilon$ compared to those that are initially not in any resonance. This is because in the initially resonant systems the interactions first need to break the resonance before $\epsilon$ can grow freely. Interestingly, if negative, $\Delta\epsilon$ remains small, $\left|\Delta\epsilon\right|\lesssim10^{-2}$. Thus, while some systems may exhibit $\Delta\epsilon<0$, they do not move significantly away from their initial offset. In contrast, in case of $\Delta\epsilon>0$, the planetary orbits can diverge much more significantly.

\subsubsection{Initially non-resonant pairs}
\label{sec:epsilon_nonres}
\begin{figure}[htb]
\plotone{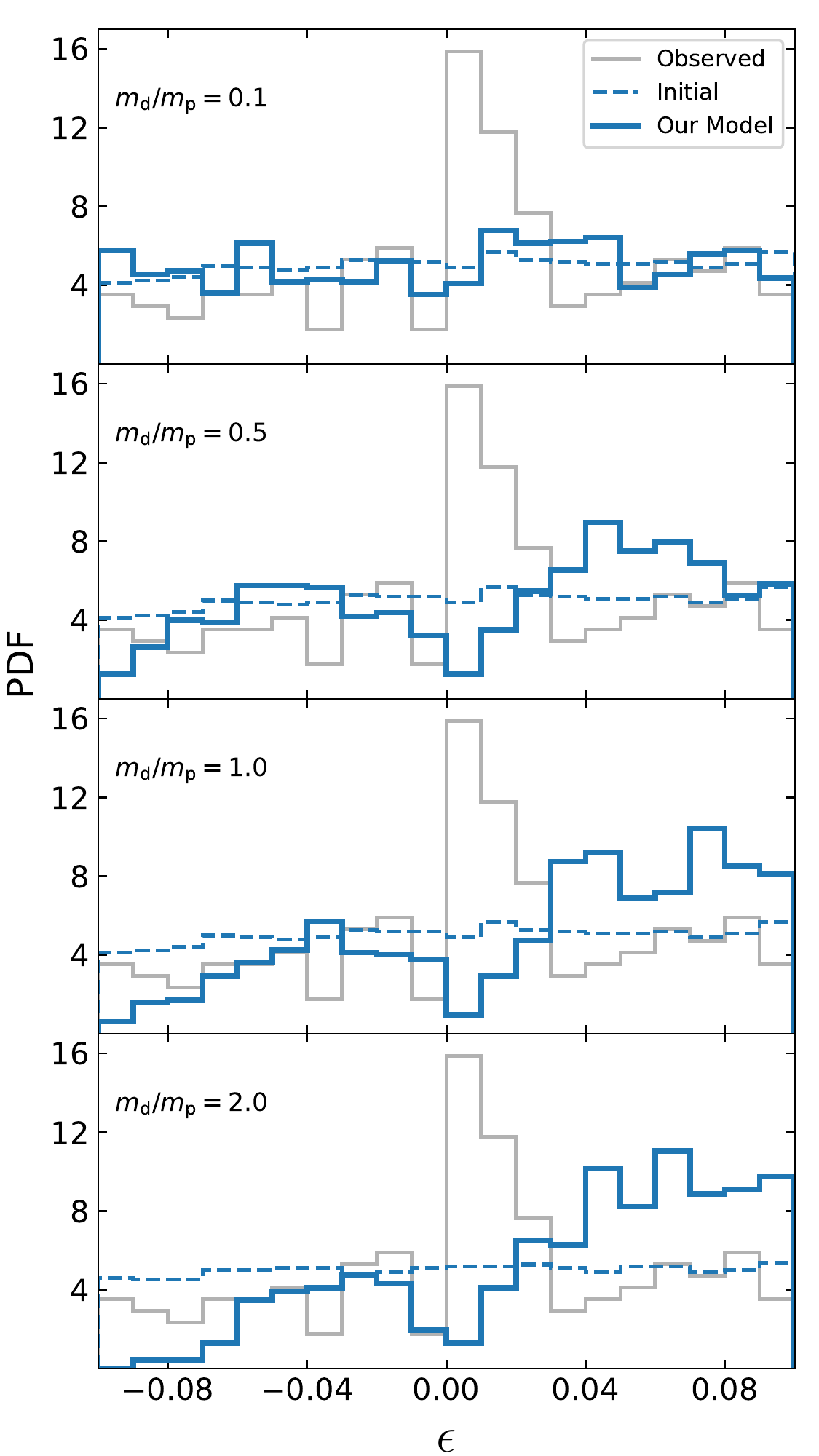}
\caption{Distribution of $\epsilon$ for initially non-resonant systems near $3:2$. Dashed (blue) and solid (blue) denote the initial ($\epsilonini$) and final ($\epsilonfin$) distributions, respectively. Different panels show results from different initial $\mdisk/\mplanet$ (see legend). Grey (solid) denotes the $\epsilon$ distribution of observed systems as a reference. Starting from a flat distribution, planetesimal interactions naturally produce a deficit of planet pairs near $\epsilon=0$ for a sufficiently massive planetesimal disk ($\mdisk/\mplanet\geq0.5$). The peak wide of $\epsilon=0$ in the observed distribution is not reproduced.
\label{fig:eps_nr_32}}
\end{figure}
\begin{figure}[htb]
\plotone{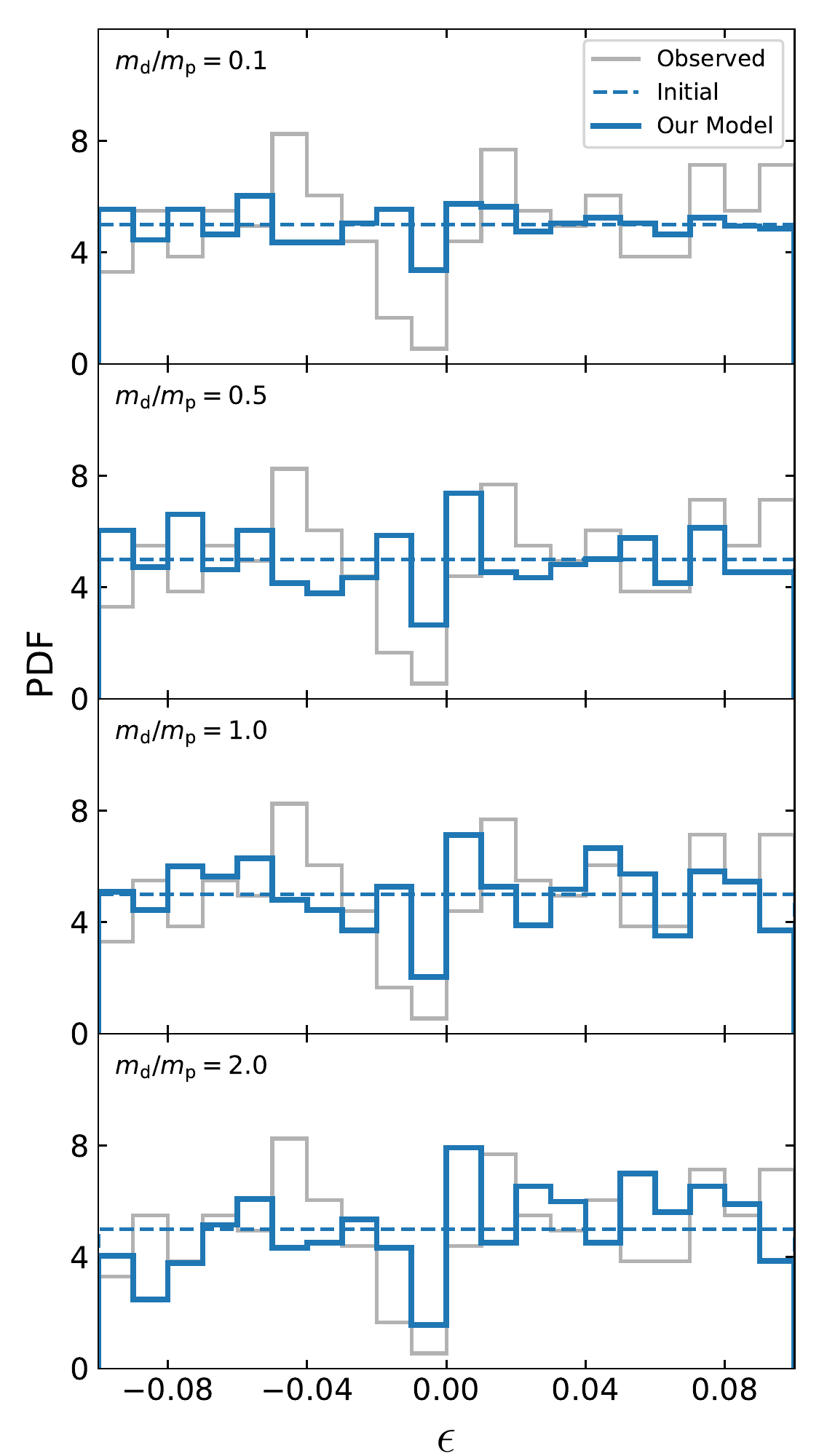}
\caption{Same as \autoref{fig:eps_nr_32} but for non-resonant planet pairs near $2:1$. The deficit of simulated planet pairs narrow of $\epsilon = 0$ is prominent for initial $\mdisk/\mplanet\geq0.5$.}
\label{fig:eps_nr_21}
\end{figure}
\autoref{fig:eps_nr_32} and \autoref{fig:eps_nr_21} show the $\epsilon_{\rm{fin}}$ distribution of initially non-resonant planet pairs near the $3:2$ and $2:1$ MMRs for models with different initial $\mdisk/\mplanet$. The blue solid (dashed) histograms show the final (initial) $\epsilon$ distributions. We show the observed distribution in grey for reference. Interactions with the planetesimal disk create a clear deficit of planet pairs near the nominal positions for both resonances. The deficits become particularly striking if compared with the adopted flat initial distribution of $\epsilon$ in our models. As expected, in both cases, the deficit is more prominent in models with higher $\mdisk/\mplanet$.  

The origin of the deficit is not hard to understand. Planet pairs, initially narrow or wide of a resonance, that do not cross $\epsilon=0$, exhibit a smooth increase in $\epsilon$ over time due to interactions with the planetesimals (e.g., \autoref{fig:evo_nr}).This smooth increase in $\epsilon$ gradually shifts planet pairs from lower to higher $\epsilon$. In contrast, when planet pairs cross $\epsilon=0$, $\epsilon$ increases rapidly (e.g., \autoref{fig:eps_evo_nr_cross}). This sudden jump in $\epsilon$ creates the deficit just narrow of the resonance, while the width of the deficit is dependent on the magnitude of this jump.

For any given $\mdisk/\mplanet$, the deficit near $3:2$ is wider than that near $2:1$. In contrast, for any given $\mdisk/\mplanet$, the deficit narrow of $2:1$ is more prominent compared to that narrow of $3:2$. These differences can be understood by noticing the difference in $\Delta\epsilon$ magnitudes near $3:2$ and $2:1$, for any given $\mdisk/\mplanet$ (\autoref{fig:del_eps_comb}). Narrow of a resonance, there is a competition between how efficiently planet pairs can cross $\epsilon=0$ creating a deficit, and replenish this deficit from lower period ratios via smooth $\epsilon$ increase. Since $\Delta\epsilon$ is statistically larger for planet pairs near $3:2$ (compared to that for planet pairs near $2:1$) for any given $\mdisk/\mplanet$, replenishment of the deficit is more efficient narrow of $3:2$. Interestingly, in the observed systems too, the deficit narrow of $2:1$ is statistically more significant compared to that narrow of $3:2$ \citep{2015Steffen}.

For the same reason, the deficit wide of $3:2$ extends to larger positive $\epsilon$ compared to $2:1$. For example, for $\mdisk/\mplanet\geq0.5$, the distribution of model systems with $\epsilon>0$ near $3:2$ peaks at a higher $\epsilon$ compared to the $\epsilon$ corresponding to the peak in the observed systems. This suggests that planetesimal interactions with only initially non-resonant planet pairs may not fully explain the deficit as well as the excess of systems observed narrow and wide of $3:2$. In contrast, while the bins narrow of $2:1$ show a clear deficit of model systems, those wide of $2:1$ do not exhibit any significant deficits.

\subsubsection{Initially resonant pairs}
\label{sec:epsilon_res}

\begin{figure}[htb]
\plotone{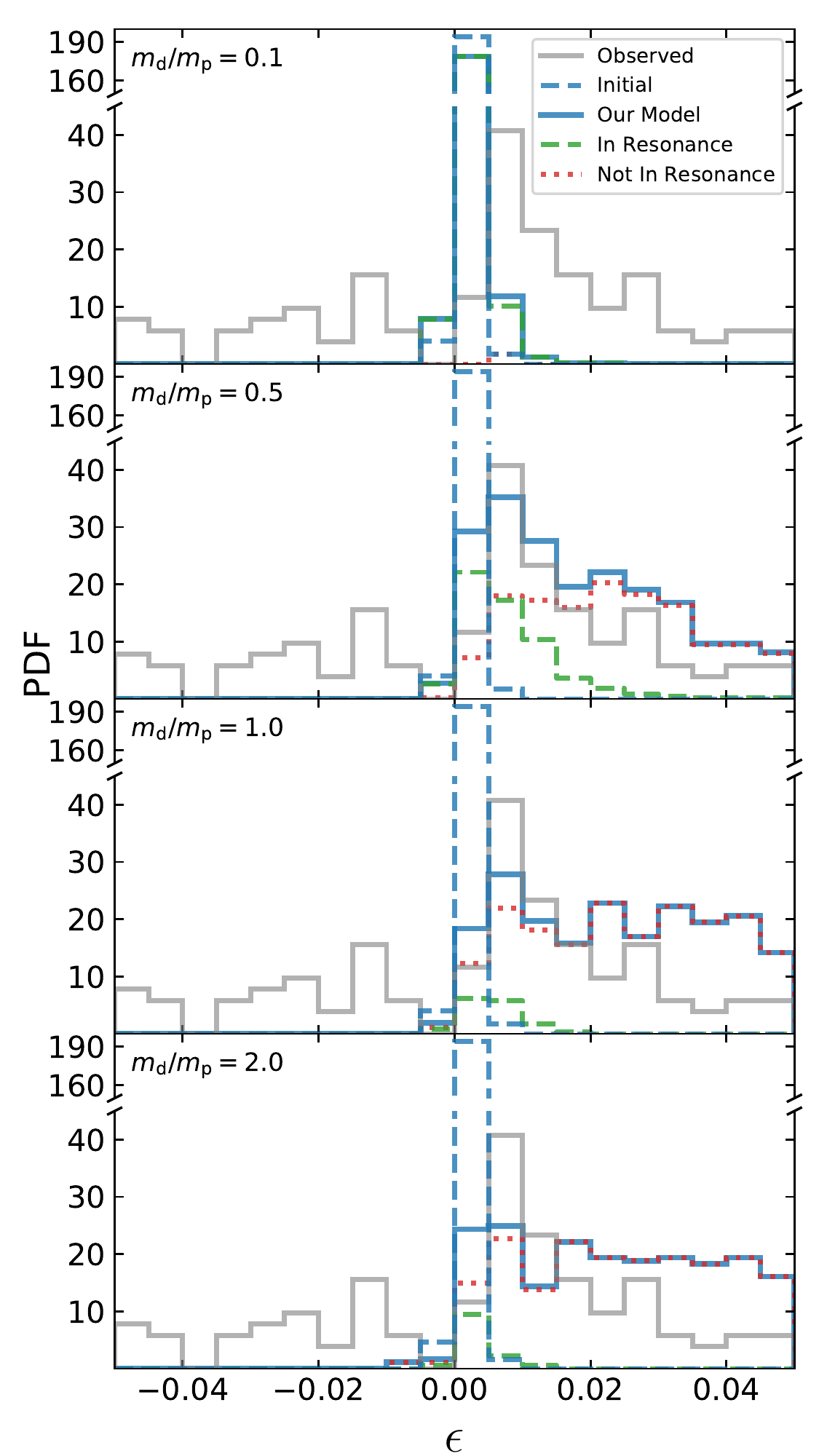}
\caption{Same as \autoref{fig:eps_nr_32} but for model systems initially trapped in the $3:2$ MMR. Blue solid (dashed) shows the final (initial) $\epsilon$ distributions of the modeled systems. Solid grey shows that for the observed systems for reference. Green dashed (red dotted) shows the $\epsilonfin$ distribution of modeled systems that remain in resonance (break out of the resonance). As $\mdisk/\mplanet$ increases increasing $\minteract/\mplanet$, higher fractions of systems break out of resonance. Systems that break out of resonance exhibit significantly higher positive $\epsilon$ relative to those that do not (\autoref{tab:reso_break}). The $\epsilon$ distributions for simulated and observed systems exhibit peaks at similar positive $\epsilon$ for models with initial $\mdisk/\mplanet\geq0.5$.
\label{fig:eps_r_32}}
\end{figure}
\begin{figure}[htb]
\plotone{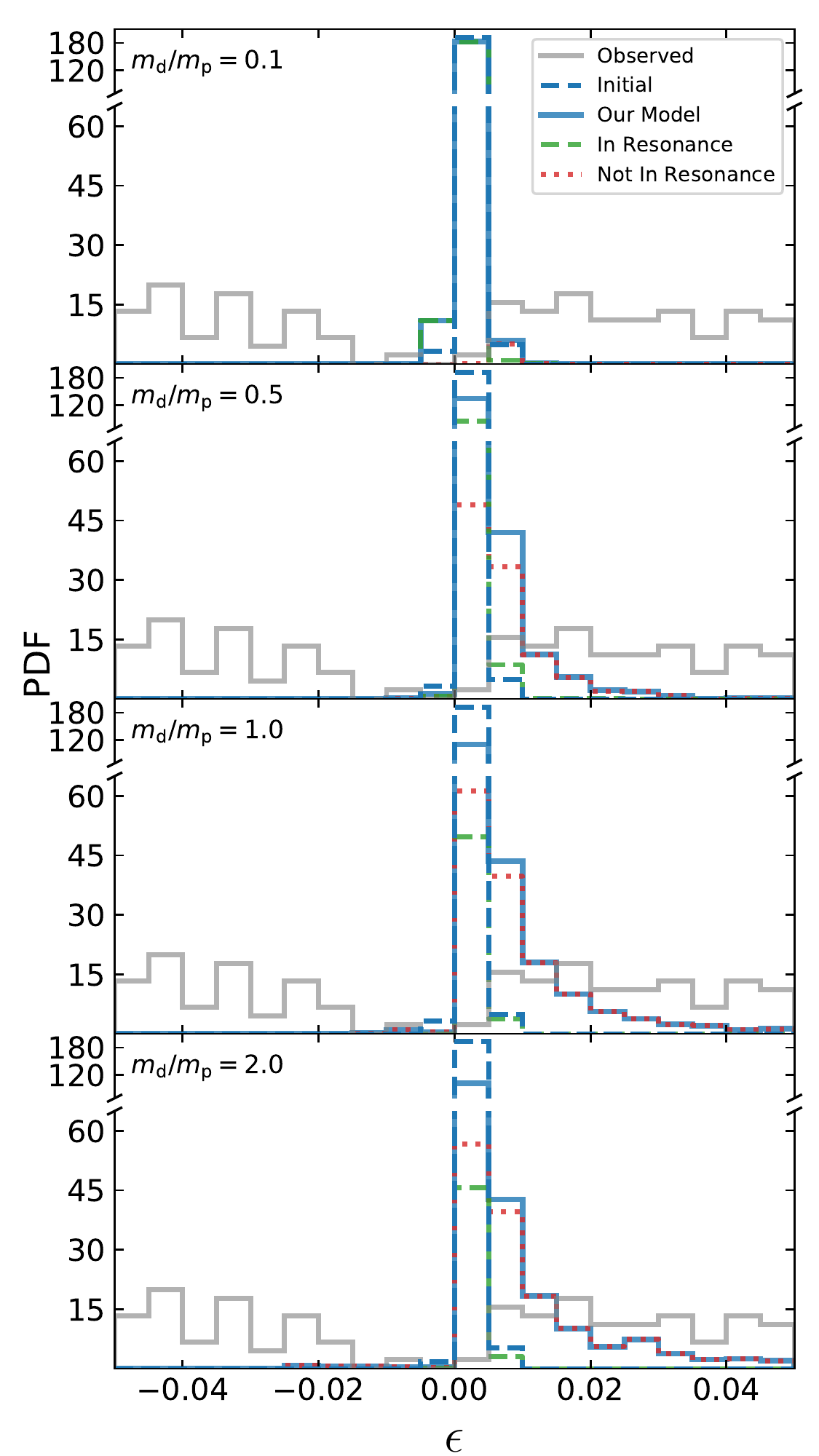}
\caption{Same as \autoref{fig:eps_r_32} but systems initially in the $2:1$ MMR. A higher fraction of systems remain in resonance till the simulation stopping time compared to systems initially in the $3:2$ MMR for any given initial $\mdisk/\mplanet$ (also see \autoref{tab:reso_break}).
\label{fig:eps_r_21}}
\end{figure}
\autoref{fig:eps_r_32} shows the initial and final distributions for $\epsilon$ for models where planet pairs are initially trapped in the $3:2$ MMR. Initially, all planet-pairs are concentrated in the bin just wide of $\epsilon = 0$. As a result of planet-planetesimal interactions, $\epsilon$ fluctuates and grows on an average while the planet pairs are still in resonance. If the resonance breaks, $\epsilon$ can freely increase and a monotonic increase in $\epsilon$ follows (\autoref{fig:evo_32_r}). Hence, planet pairs start populating the bins wide of the resonance. 

\begin{deluxetable*}{c|ccc|ccc}
\tablecaption{
\label{tab:reso_break}}
\tablewidth{0pt}
\tablehead{
    \colhead{$\mdisk/\mplanet$} & 
    \multicolumn{3}{c}{$3:2$ MMR} & 
    \multicolumn{3}{c}{$2:1$ MMR} \\
    \cline{2-7}
    \colhead{} & 
    \colhead{$F_{r}$} & 
    \colhead{$\epsilon_{\rm{r}} \times 10^{3}$} & 
    \colhead{$\epsilon_{\rm{nr}} \times 10^{3}$} &
    \colhead{$F_{r}$} & 
    \colhead{$\epsilon_{\rm{r}} \times 10^{3}$} &
    \colhead{$\epsilon_{\rm{nr}} \times 10^{3}$}
    }

\startdata
   0.1 &  0.99 &  $1^{+2}_{-1}$ & $6^{+.08}_{-.08}$ 
   &  0.97 &  $1^{+.6}_{-.3}$ & $10^{+.06}_{-.06}$ \\
   0.5 &  0.27 &  $6^{+8}_{-4}$ & $26^{+21}_{-16}$
   &  0.48 &  $2^{+2}_{-1}$ & $5^{+6}_{-3}$ \\
   1.0 &  0.04 &  $5^{+3}_{-4}$ & $45^{+49}_{-30}$
   &  0.26 &  $2^{+2}_{-1}$ & $6^{+11}_{-4}$ \\
   2.0 &  0.02 &  $2^{+4}_{-2}$ & $96^{+95}_{-66}$
   &  0.23 &  $2^{+2}_{-1}$ & $8^{+20}_{-5}$ \\
\enddata

\tablecomments{Fraction of systems ($F_{r}$) where the initial MMR is not broken by the integration stopping time for the two resonances we consider for different initial $\mdisk/\mplanet$.The median value and $1\sigma$ range of $\epsilonfin$ are given for systems that are still in resonance ($\epsilon_{\rm{r}}$) and those that broke out of resonance ($\epsilon_{\rm{nr}}$).} 
\end{deluxetable*}

As expected, higher-mass disks lead to higher levels of perturbations by the planetesimals and as a result, resonance breaks in a higher fraction of systems, and the planet pairs are pushed to higher $\epsilon$ values. In a significant fraction of our simulated initially resonant systems, especially those with the lower $\mdisk/\mplanet$, the planet-planetesimal interactions are not sufficient to break the resonance. For example, $\approx99\%$ ($\approx2\%$) of our models initially trapped in the $3:2$ MMR embedded in an initial disk with $\mdisk/\mplanet=0.1$ ($2$) remain trapped in resonance at the integration stopping time (see \autoref{tab:reso_break}). 

\autoref{fig:eps_r_21} shows the $\epsilon$ distributions for systems that are initially trapped in the $2:1$ MMR. While the qualitative nature in this case is very similar to that for systems initially trapped in $3:2$ MMR, a notable difference is that it is harder to break the $2:1$ MMR (\autoref{tab:reso_break}). For example, even in our models with the highest $\mdisk/\mplanet$, $23\%$ systems remain trapped in $2:1$ MMR  (in contrast, only $2\%$ remain trapped in $3:2$). As a result, wide of the $2:1$ MMR, the $\epsilon$ distribution does not exhibit a peak, but rather a steady decrease from the peak corresponding to the systems still trapped in resonance. The range in final $\epsilon$ for systems initially in the $3:2$ MMR is much larger compared to those initially in $2:1$.

Interestingly, we find excellent agreement between the locations of the peaks in the observed and model $\epsilon$-distributions for systems initially in the $3:2$ MMR, especially for models using $\mdisk/\mplanet =0.5$, $1$.

\subsection{Comparison with observed systems} \label{sec:resmix}
So far we focused on describing the effects of planetesimal scattering for ensembles that completely consist of either initially non-resonant or initially resonant systems. However, reality almost certainly is not as simple and it is expected that a fraction of the observed systems near $3:2$ or $2:1$ may have been trapped in MMR in the past. In addition, the amount of perturbations from a planetesimal disk a planetary system experiences also likely varies from system to system. Hence, to compare our simulated systems with those observed, we need to combine initially resonant and non-resonant systems that are allowed to go through varying degrees of planetesimal perturbations. In what follows, we first describe how we combine all our simulated models, then we compare the simulated and observed $\epsilon$ distributions near $3:2$ and $2:1$ MMRs.  

\subsubsection{Combining models}
\label{sec:mcmc}

We have already seen that $\Delta\epsilon$ is correlated with $\Delta\mplanet/\mplanet$, our adopted proxy for the level of perturbation provided by the planetesimal disk (\autoref{sec:epsilon}, \autoref{fig:del_eps_comb}). Furthermore, a-priori we do not know the relative contributions of initially resonant and non-resonant systems, $\beta\equiv\nres/(\nres + \nnonres)$, for the observed planet pairs. Here, $\nres$ ($\nnonres$) denotes the number of initially resonant (non-resonant) systems and $\nres+\nnonres=\ntot$, the total number of planet pairs. We treat $\Delta\mplanet/\mplanet$ and $\beta$ as parameters and using our models we estimate the posterior distributions of $\Delta\mplanet/\mplanet$ and $\beta$ given the observed $\epsilon$ distributions using Markov-chain Monte Carlo (MCMC). \footnote{Note that instead of $\Delta\mplanet/\mplanet$, in principle, one can use other measures of planet-planetesimal interactions including $\minteract/\mplanet$. However, it is simply easier to track $\Delta\mplanet/\mplanet$ in our simulations.}

We use \emcee\ \citep{2013emcee} to run our MCMC for a two-dimensional parameter estimation. We find the posteriors separately for systems near $2:1$ and $3:2$. For systems near each resonance, we combine all models for all initial $\mdisk/\mplanet$ into two separate sets, initially resonant and initially non-resonant. For any given value of $\Delta\mplanet/\mplanet$ we select systems that are within $\Delta\mplanet/\mplanet\pm0.02$ separately from the initially resonant and non-resonant sets. Then, for any given value of $\beta$, we randomly select (with replacement) $\nres$ ($\nnonres$) systems from these subsets (selected based on $\Delta\mplanet/\mplanet$) from the initially resonant (non-resonant) systems such that $\nres/\ntot=\beta$ and $\ntot=\nres+\nnonres=10^3$. We create a PDF for $\epsilon$ based on this randomly selected mixture of planet pairs with specific $\{\beta, \Delta\mplanet/\mplanet\}$ using kernel densities with Gaussian kernels. Note that the observed number of pairs we want to compare with is fewer in number by a factor of $\sim 10$. We intentionally choose a high sample size from our models to reduce statistical noise in the model PDF for $\epsilon$. We define the log-likelihood of a particular value $\left\{ \beta, \frac{\Delta\mplanet}{\mplanet} \right\}_j$ for systems near any given resonance as-
\begin{equation}
	\ln\mathcal{L}\left[ \left\{ \beta, \frac{\Delta\mplanet}{\mplanet} \right\}_j \right] = \sum_i \ln\mathcal{P}\left[ \epsilon_{\rm{obs},i}\ |\ \left\{\beta, \frac{\Delta\mplanet}{\mplanet} \right\}_j \right]
\end{equation}
where, the index $i$ denotes the observed systems, $\mathcal{P}$ denotes the conditional probability of the occurrence of $\epsilon_{\rm{obs},i}$ given $\left\{\beta, \frac{\Delta\mplanet}{\mplanet}\right\}_j$. We then proceed to find the posterior distribution, $\mathcal{P} \left( \left\{\beta, \frac{\Delta\mplanet}{\mplanet}\right\}\ |\ \vec{\epsilonobs}\right)$, where, $\vec{\epsilonobs}$ is the observed distribution of $\epsilon$ for near-resonant systems corresponding to either $P_2/P_1= 3:2$ or $2:1$. For each value of $P_2/P_1$, we use $256$ walkers and $20,000$ steps for each walker. We use the default \texttt{StretchMove} method in the \emcee\ package for the jumps. We discard the first $5000$ steps as burn in. Furthermore, we prune the walker positions by choosing $1$ position in $10$. We use flat priors, i.e., $\beta$ is flat between $0$ and $1$ and $\Delta\mplanet/\mplanet$ is flat between $0$ and $30\%$ based on range we find in our simulations (e.g., \autoref{fig:del_eps_comb}).

\subsubsection{Estimated parameters}
\label{sec:bestfit_params}
\begin{figure*}[htb]
\plotone{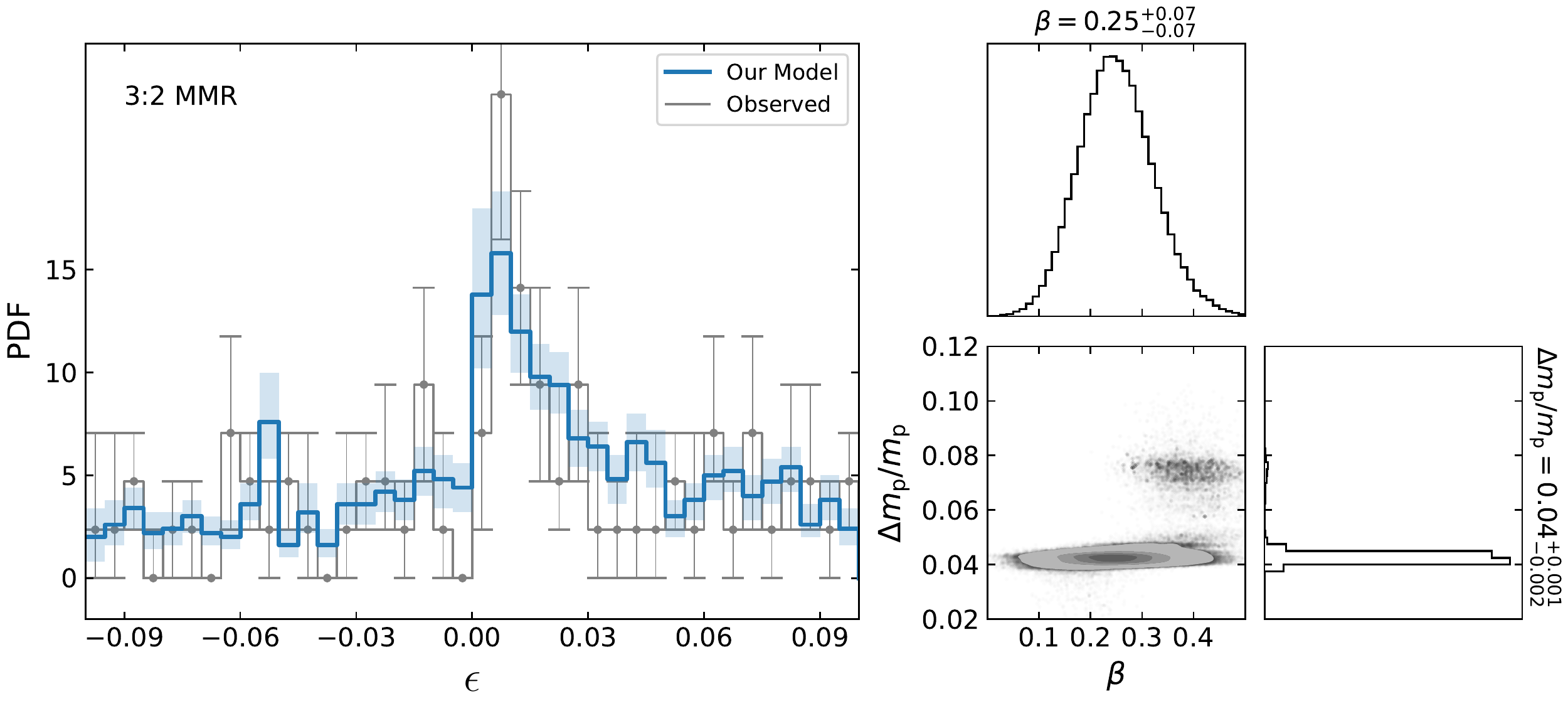}
\caption{{\em Left}: The $\epsilon$ distributions for the best fit mixed model (blue) and the observed systems (grey) near $3:2$. Blue shades and grey error bars denote $1\sigma$ spreads in the simulated and observed distributions, respectively. We find excellent agreement with the observed $\epsilon$ distribution; the simulated and observed distributions overlap within $1\sigma$ for almost all bins. The prominent observed peak wide of $3:2$ is clearly reproduced in the simulated distribution. {\em Right}: Corner plot showing the posterior distributions for $\fracdelm$ and $\beta$. The median and $1\sigma$ confidence intervals are also shown for both parameters. The posterior distribution of $\beta$ is quite broad and excludes zero indicating a clear preference for a non-zero contribution from systems initially trapped in the $3:2$ MMR.
\label{fig:grand_mix_2d_32}}
\end{figure*}
\begin{figure*}[htb]
\plotone{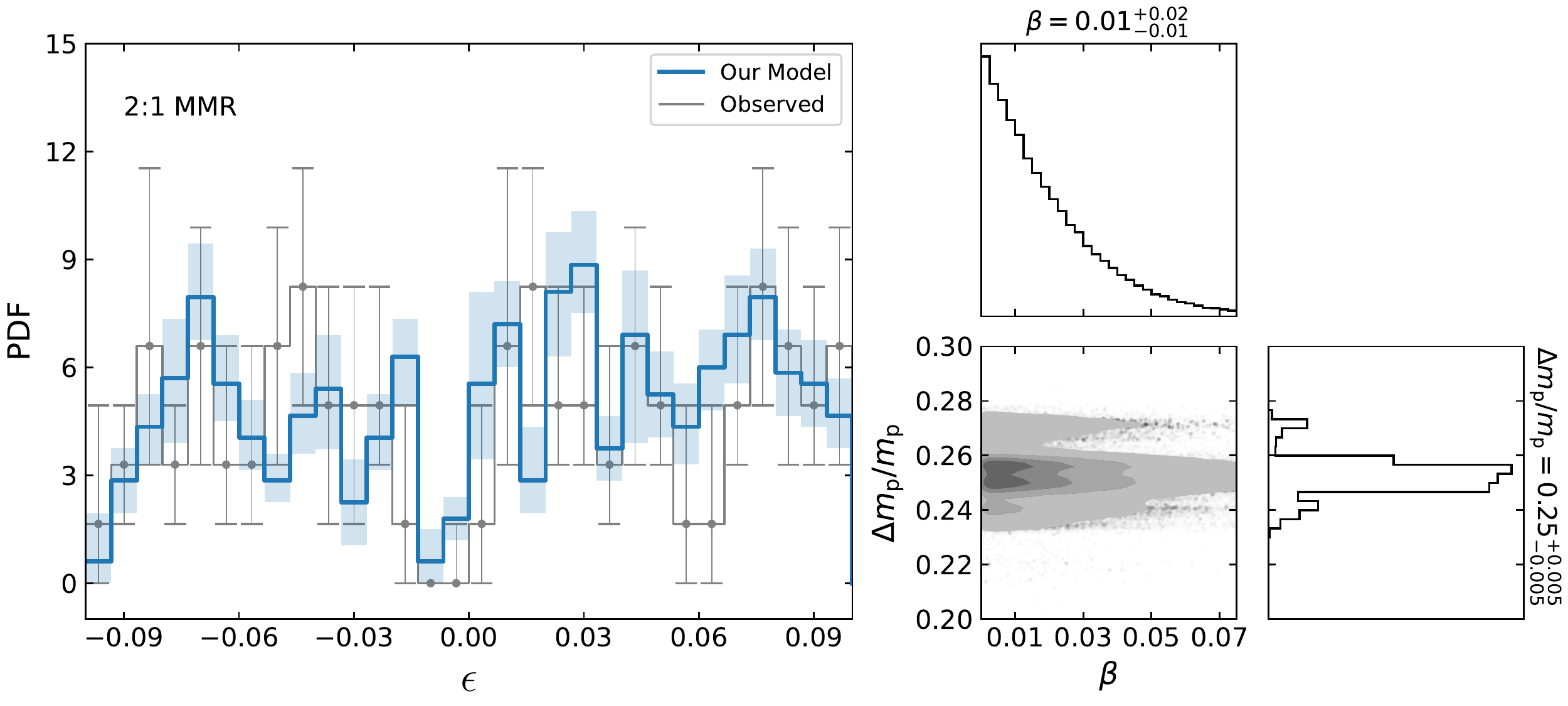}
\caption{Same as \autoref{fig:grand_mix_2d_32} but for systems near $2:1$. We find excellent agreement between the $\epsilon$ distributions of simulated and observed systems including the deficit narrow of $2:1$. The posterior distribution of $\beta$ monotonically increases all the way down to $\beta=0$, indicating that there is little contribution from systems initially in the $2:1$ MMR. The median of the posterior distribution for $\fracdelm\approx 0.25$, which is higher than the same found for systems near $3:2$ MMR. This is consistent with our finding that a higher level of planetesimal interactions is needed to significantly alter planet pairs near the $2:1$ MMR (e.g., \autoref{fig:del_eps_comb}).
\label{fig:grand_mix_2d_21}}
\end{figure*}
\autoref{fig:grand_mix_2d_32} shows the posterior distributions for $\fracdelm$ and $\beta$ as well as the $\epsilon$ distributions created using the parameter values drawn from the posteriors for simulated systems near $3:2$. The corresponding observed distributions are also shown for comparison. We use bootstrap by randomly selecting (with replacement) half the number of observed planet pairs $10^4$ times and create distributions for all of them. The histogram heights and the error bars for the observed distribution in the figure show the median and the span between the 16th and 84th percentiles for each bin from our bootstrap exercise. We generate the simulated histogram in the following way. We randomly draw $10^4$ samples of $\left\{\beta, \fracdelm\right\}$ from the posterior distribution. For each of these draws, we create a mixed set of planet pairs and collect the values of $\epsilon$. Using these $10^4$ synthetic distributions, we create the $\epsilon$ distribution where the histogram heights and the shaded regions denote the median and the $16$th and $84$th percentiles in each bin. 

Overall, the model and observed distributions of $\epsilon$ for planet pairs near $3:2$ agree within their respective $1\sigma$-equivalent errors. The bin immediately narrow of $3:2$ is the only one where the simulated and the observed distributions are apparently not within $1\sigma$-equivalent errorbars of each other. The errorbar for this bin in the observed distribution is likely grossly underestimated. Since in bootstrap, any bin that contain zero systems, would always contain zero systems in all bootstrap draws, and hence, would show zero errorbar, which is not realistic. In addition, as discussed earlier, the prominence of the deficit narrow of a resonance is dependent on the competition between the efficiency of jumping across $\epsilon=0$ and that of replenishment due to smooth increase in $\epsilon$ (\autoref{sec:epsilon_nonres}). The lack of prominence of the deficit in the bin immediately narrow of $3:2$ for the simulated systems may be an artifact of our assumed flat initial $\epsilon$ distribution. A close inspection of the observed systems indicates that in reality, the $\epsilon$ distribution narrow of $3:2$ has a decreasing trend as we move away narrow from $3:2$. Due to this difference, we may be replenishing the deficit in the bin immediately narrow of $3:2$ more in our simulations compared to the observed systems.

For systems near $3:2$, although the $\beta$ posterior distribution is broad, there is a clear preference for a non-zero contribution from systems initially trapped in a $3:2$ MMR. Our models suggest that planetesimal interactions can naturally create the deficit narrow of $3:2$ from an ensemble of non-resonant planet pairs with an initially flat $\epsilon$ distribution (\autoref{sec:epsilon_nonres}), while a non-zero contribution of systems initially trapped in the $3:2$ MMR, via planetesimal interactions, helps create the excess of systems just wide of $3:2$, as observed. For the systems near $3:2$, the dominant peak in the posterior distribution is approximately at $\left\{\beta,\fracdelm\right\} = \left\{0.25, 0.04\right\}$. Although contribution from initially resonant systems are needed to explain the excess of observed systems wide of $3:2$, most of these initially resonant systems ($\sim 84 \%$) are not in resonance in the end. We find that only $4^{+2}_{-1} \%$ of systems in the range $-0.1<\epsilon<0.1$ are finally in resonance. These resonant systems predominantly occupy the bin right next to the nominal MMR at $\epsilon=0$. In contrast, the second bin wide of the MMR is the most populated in the observed $\epsilon$ distribution (\autoref{fig:grand_mix_2d_32}).

The situation around $2:1$ is different. For systems near $2:1$ the $\beta$ posterior distribution monotonically increases all the way to $\beta=0$ indicating that the observed $\epsilon$ distribution prefers little contribution from systems initially trapped in the $2:1$ MMR (\autoref{fig:grand_mix_2d_21}). Our results suggest that this is because breaking the $2:1$ MMR is considerably harder compared to $3:2$. Any significant contribution from planet pairs initially trapped in the $2:1$ MMR creates a $\epsilon_{\rm{fin}}$ distribution exhibiting an excess too close to $2:1$ compared to what is observed. However, the deficit narrow of $2:1$ can be naturally explained as a result of planet-planetesimal interactions of initially non-resonant planet pairs.

The relative ease in breaking the resonance between the $3:2$ and $2:1$ also creates a significant difference between where the posteriors for $\fracdelm$ peak for $3:2$ and $2:1$; while, the primary peak in the posterior distribution for $3:2$ is near $\fracdelm\approx 0.04$, for $2:1$ it is near $\fracdelm\approx 0.25$. Clearly, a higher level of planetesimal interactions is necessary to significantly perturb the planet pairs near or initially trapped in the $2:1$ MMR.

\section{Summary and Discussion} \label{sec:discuss}

In this study, we have investigated the evolution of planet pairs in or in the vicinity of $3:2$ and $2:1$ MMRs as a result of interactions with nearby planetesimals in a residual disk after gas dispersal. Using a large number of $N$-body simulations involving two planets and thousands of planetesimals, we have investigated whether such interactions, expected to be common immediately after gas dispersal, can explain the observed deficit and excess of planet pairs narrow and wide of these period ratios. 

We find that interactions with planetesimals typically increase the period ratios of planet pairs, trapped in an MMR or not (\autoref{fig:evo_nr}). For a small fraction of systems the period ratios may decrease, however, in these cases, the fractional change is much smaller compared to that when the period ratios increase (e.g., \autoref{fig:del_eps_comb}). If the increase in period ratios for planet pairs that are initially narrow of an MMR brings them sufficiently close to the MMR ($\epsilon \gtrsim -0.005$), the planet pairs overshoot the resonance and get deposited wide of $\epsilon=0$ (\autoref{fig:eps_evo_nr_cross}). Following the jump, the offset keeps on increasing as long as the planets have access to planetesimals to interact with. 

Using an initially flat distribution of $\epsilon$ across $3:2$ and $2:1$, we find that planetesimal interactions naturally create a deficit of systems narrow of the nominal resonance positions at $\epsilon=0$. For systems near the $2:1$ MMR, the final $\epsilon$ distribution generated from initially non-resonant planet pairs closely resembles that for the observed adjacent planet pairs for any disk with $\mdisk/\mplanet\geq 0.5$ (\autoref{fig:eps_nr_21}). However, for model systems near the  $3:2$ MMR, the deficit spans a wider range of $\epsilon$ and can extend beyond where actually an excess is observed (\autoref{fig:eps_nr_32}). This indicates that while the observed deficit of systems narrow of $3:2$ can be explained by the evolution of initially non-resonant planet pairs embedded in a disk of residual solids, the observed excess wide of $3:2$ may not be explained using initially non-resonant planets only.

Our models with initially resonant planet pairs show that the offset typically remains limited to $\epsilon\sim10^{-3}$, while the planet pairs are still in resonance. However, once the random fluctuations break the resonance, $\epsilon$ monotonically increases (\autoref{fig:evo_32_r}).  
For any given $\mdisk/\mplanet$ it is significantly easier to break the $3:2$ MMR compared to breaking the $2:1$ MMR (e.g., \autoref{fig:eps_r_32}, \autoref{fig:eps_r_21}, \autoref{tab:reso_break}). 

Using the simulated ensembles of initially resonant and non-resonant planet pairs and the observed $\epsilon$ distribution we constrain the level of planetesimal interactions ($\fracdelm$) and the relative contribution of initially resonant systems ($\beta$) to explain the observed asymmetric abundances across $3:2$ and $2:1$. We find that a significant fraction ($\beta \approx 0.25$) of initially-resonant planet pairs is needed to explain the observed $\epsilon$ distribution across $3:2$. While the deficit narrow of $3:2$ can be explained by planetesimal perturbations on initially-non-resonant planet pairs, the observed excess wide of 3:2 requires contribution from initially-resonant pairs. Interestingly, the posterior distribution clearly excludes $\beta=0$ for systems near $3:2$. For these systems, $\fracdelm\approx4\%$ is most preferred (\autoref{fig:grand_mix_2d_32}). Higher $\Delta\mplanet/\mplanet$ would create larger $\Delta\epsilon$ and the excess wide of $3:2$ would move further away compared to the location where it is observed. Although, a significant fraction of initially resonant planet pairs are needed to explain the observed $\epsilon$ distribution across $3:2$, only about $4\%$ of all final systems are expected to be in resonance within $-0.1<\epsilon<0.1$. This finding is broadly consistent with the findings of \citet{2017Izidoro}, where they tried to explain the observed period ratio distribution from super-earth systems in compact resonant chains that undergo dynamical instability once the gas dissipates. Their simulations match the observed period ratio distribution if less than $25\%$ of resonant chains remain stable while the rest gets destabilized. Although the physical process and the initial simulation setup are quite different, interestingly, we also find that a similar fraction of initially resonant systems are is needed to explain the observed asymmetry across $3:2$. From the abundance of planet pairs wide of the MMRs they also suggested that $\sim 5 \%$ of planet pairs may be in resonance at present, very similar to the fraction of planet pairs that are still in resonance in our best fit model.

In contrast to our findings for systems near $3:2$, the observed $\epsilon$ distribution across $2:1$ supports little contribution from initially resonant planet pairs (\autoref{fig:grand_mix_2d_21}). Interestingly, this is in contrast to the implied expectations of CF15 in the details. CF15 considered only planet pairs initially in the $2:1$ MMR and showed that perturbations from a disk of solids can break the resonance and preferentially increase $\epsilon$. While the evolution of embedded planets pairs found in this study is in qualitative agreement with the findings of CF15, our results suggest that a significant contribution from planet pairs initially in $2:1$ MMR would be inconsistent with the observations. Our simulations also indicate that a higher level of planetesimal interactions ($\fracdelm\approx25\%$) is needed to explain the observed $\epsilon$ distribution across $2:1$.

The difference in the relative contributions from initially resonant planet pairs in the $3:2$ and $2:1$ MMRs found in this study is interesting and is likely related to the types of planets we study here. For example, \citet{Deck2015} suggested that due to type-I migration, relevant for the planets we study here, the stability of the $2:1$ resonance requires a more demanding planet-planet mass ratio, inner planet $12\times$ more massive than the outer planet, compared to $3:2$, where the required mass ratio is $\sim 1$. This is consistent with our requirement of a higher fraction of systems trapped into the $3:2$ MMR compared to that trapped into the $2:1$ MMR at the onset of planetesimal-driven evolution.

Overall, we find excellent agreement between the $\epsilon$ distributions of our simulated and observed systems across both $3:2$ and $2:1$. The excellent agreement between models and observations with little need for fine-tuning indicates that planetesimal interactions can explain the $\epsilon$ distribution across $3:2$ and $2:1$. Our models also indicate that up to a few percent of all near-resonant systems currently may be in resonance. While our finding of low fraction of true resonances at present are consistent with some past studies \citep[e.g.,][]{Veras2012, Deck2015, 2017Izidoro, 2021_Izidoro}, it is in contradiction with proposed scenarios that invoke damping while still in resonance \citep[e.g.,][]{2020Choksi}. Future measurements of the relative abundance of truly-resonant pairs among the near-resonant can ultimately shed light on the allowed mechanisms out of the many proposed.

\begin{acknowledgements}
We thank the anonymous referee for insightful comments and constructive suggestions. TG acknowledges support from TIFR's graduate fellowship. SC acknowledges support from the Department of Atomic Energy, Government of India, under project no.  12-R\&D-TFR-5.02-0200 and RTI 4002. All simulations were done using Azure cloud computing. This research has made use of the NASA Exoplanet Archive, which is operated by the California Institute of Technology, under contract with the National Aeronautics and Space Administration under the Exoplanet Exploration Program. We also thank summer student Abhijeet Singh for his initial study on this problem.
\end{acknowledgements}

\vspace{5mm}
\software{\texttt{Rebound} \citep{ReinREB2012}, \texttt{emcee} \citep{2013emcee}, \texttt{corner} \citep{corner}, \texttt{matplotlib} \citep{matplotlib_Hunter:2007}, \texttt{numpy} \citep{numpy}, \texttt{pandas} \citep{pandas, mckinney-proc-scipy-2010}, \texttt{scikit-learn} \citep{scikit-learn}}

\vspace{8mm}
\appendix
\restartappendixnumbering

\section{Total disk mass vs total mass of interacting planetesimals} \label{app:md_vs_mint}
\restartappendixnumbering
\vspace{4mm}

\begin{figure}[htb]
\plottwo{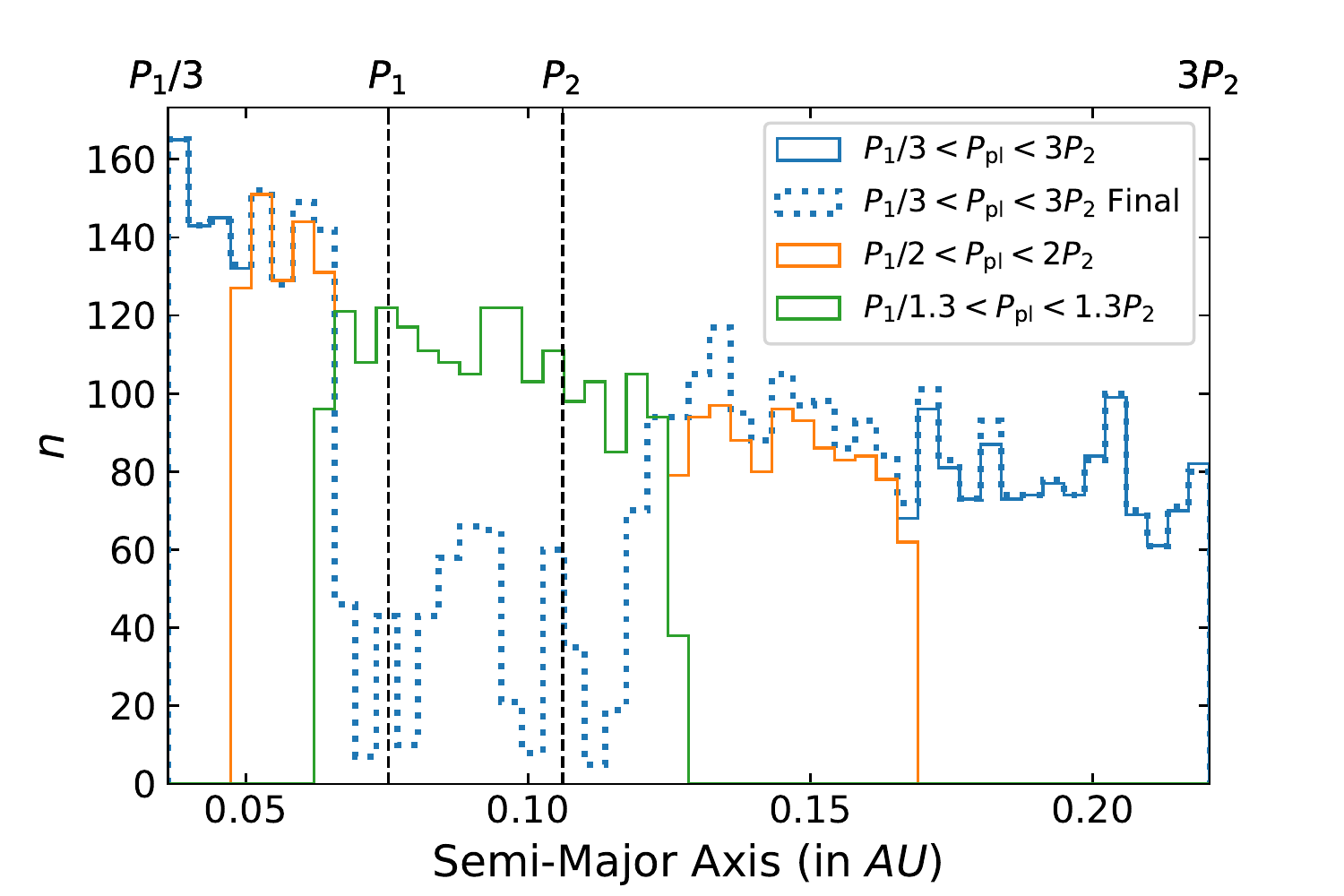}{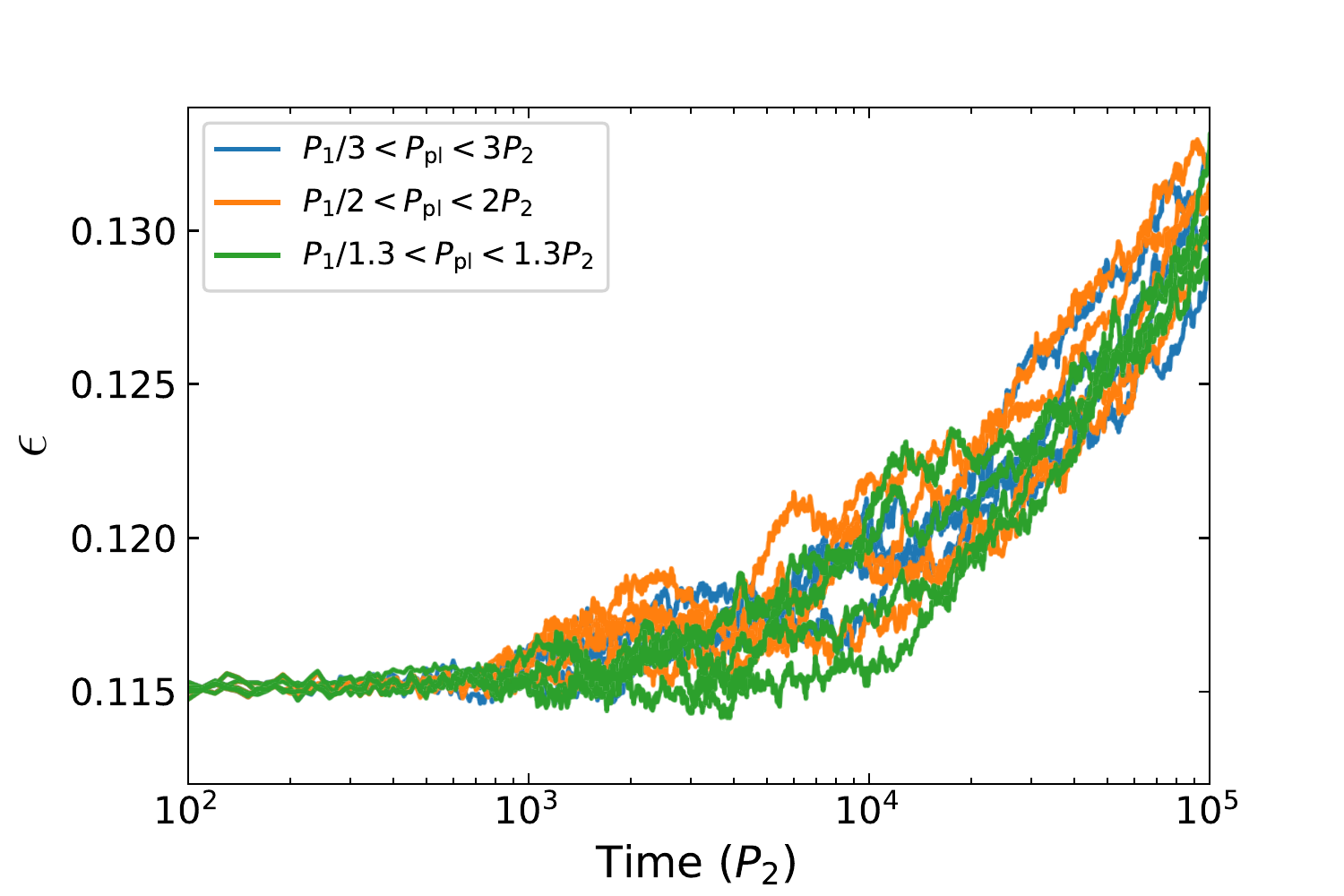}
\caption{{\em Left}: Initial planetesimal disk profiles for three different disks (solid histograms). The blue (solid line) histogram shows a disk that spans from $P_{\rm{1}}/3$ to $3P_{\rm{2}}$ and has an initial mass of $\mdisk/\mplanet = 0.5$. The distribution of this disk after Stage\ 3 is shown in the blue dotted histogram indicating the region around the planets where interacting planetesimals reside. The positions of the planets are marked by the vertical dashed lines. This example system consists of $m_1=2.1 M_{\Earth}$ and $m_2=5.48M_{\Earth}$ planets initially not in resonance around a $0.84 M_{\sun}$ star. The orange (green) histogram shows a disk that spans $P_{\rm{1}}/2<\Ppl<2P_2$ ($P_1/1.3<\Ppl<1.3P_2$) and has an initial mass of $\mdisk/\mplanet = 0.34$ ($0.19$). However, each disk has roughly the same $\minteract/\mplanet\sim0.1$ and $\Delta\mplanet/\mplanet\sim 0.09$.  
{\em Right}: $\epsilon$ evolution of the planet pair embedded in the three disks shown in the left panel. The colors denote disks denoted with the same colors in the left panel. Each line of the same color shows a different evolutionary track corresponding to a different random realisation within the same disk model. The $\epsilon$ evolution is very similar between different disks of different $\mdisk/\mplanet$ but same $\minteract/\mplanet$.}
\label{fig:diff_planetesimal_dist}
\end{figure}

The planetesimal disk we consider throughout this study is distributed in the region $P_{\rm{1}}/3 < \Ppl < 3P_{\rm{2}}$, where $\Ppl$ is the orbital period of the planetesimals. We consider this large range simply to avoid any edge effects. Although, we describe the disks via the total disk mass $\mdisk$, what really matters is the total mass of planetesimals close enough to the planets to take part in dynamical interactions, $\minteract$. This can be illustrated via the following experiment. \autoref{fig:diff_planetesimal_dist} shows the distribution of planetesimals in the same example system as shown in \autoref{fig:planetesimal_dist_stages}. In this same system, in addition to our fiducial disk spanning $P_{\rm{1}}/3 < \Ppl < 3P_{\rm{2}}$ (blue) we model the effects of two other disks truncated at closer distances from the planet pair, $P_{\rm{1}}/2 < \Ppl < 2P_{\rm{2}}$ (orange) and $P_1/1.3<\Ppl<1.3P_2$ (green). The total disk mass $\mdisk/\mplanet=0.5$, $0.34$, and $0.19$ depending on where we truncate. In each case, however, $\minteract/\mplanet=0.09$--$0.1$ and $\Delta\mplanet/\mplanet=0.087$--$0.092$ remain roughly the same. We estimate $\minteract$ by taking the difference between the planetesimal density profiles at the end of stage\ 2 and stage\ 3 within the gap carved by the planets and $\Delta\mplanet/\mplanet$ is simply the fractional change in the total planet mass. We simulate four realizations for each of the three disks simply by changing the random seed. The $\epsilon$ evolution is very similar in all three cases resulting in very similar $\epsilonfin$. The differences in $\epsilon$ between models of different disks are well within the statistical fluctuations in $\epsilon$ between different realisations of the same disk. This indicates that the evolution of the planet pair is determined by $\minteract/\mplanet$ (or equivalently, $\Delta\mplanet/\mplanet$), rather than $\mdisk/\mplanet$. 

\vspace{2cm}
\section{Individual planetesimal mass vs total mass of interacting planetesimals}\label{app:mpl_effect}
\restartappendixnumbering
\vspace{4mm}

\begin{figure}[htb]
\plotone{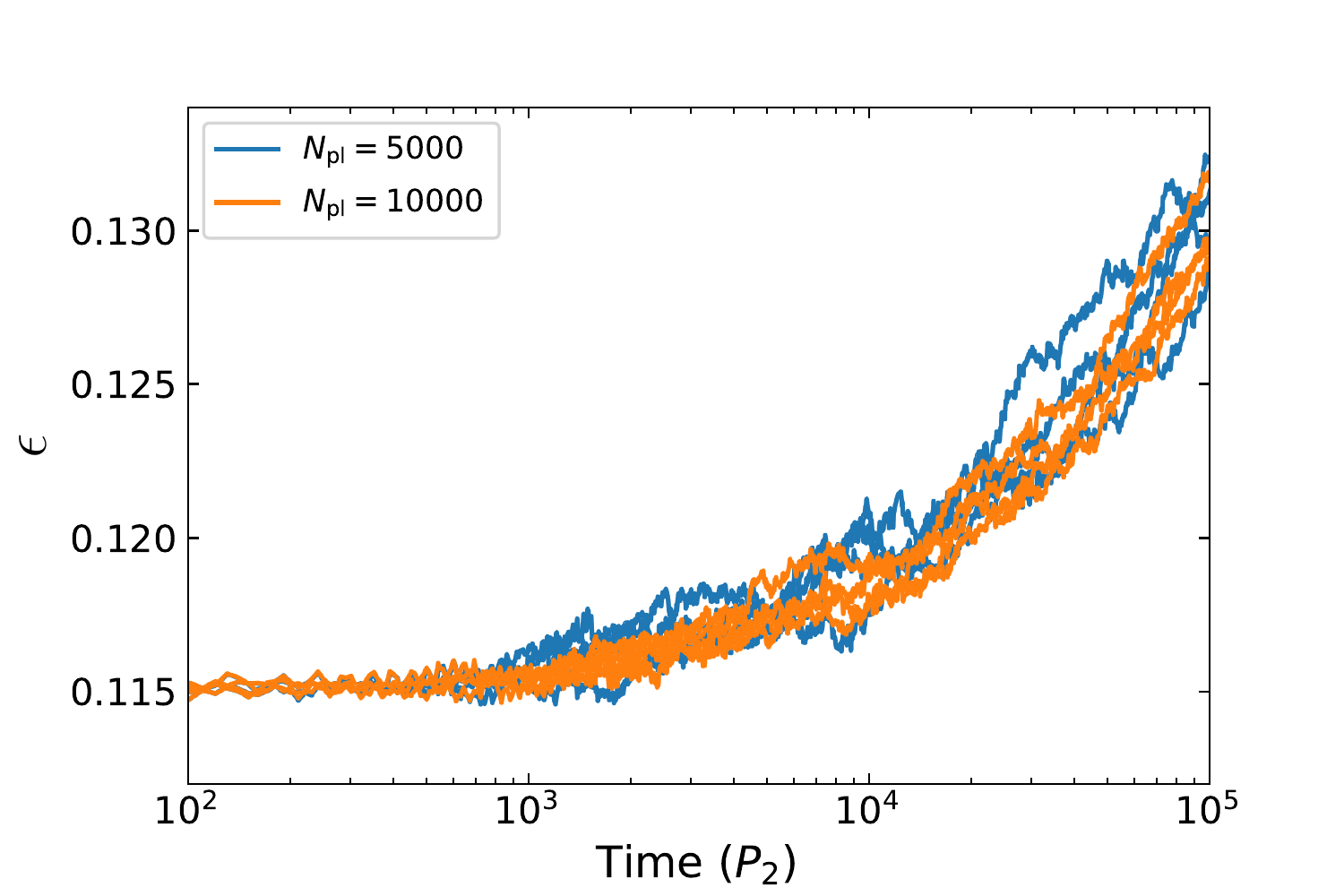}
\caption{The $\epsilon$ evolution of a planet pair embedded in planetesimal disks of equal $\mdisk/\mplanet=0.5$, but with different $\Npl=5000$ (blue) and $10000$ (orange), resulting in different individual planetesimal mass $\mpl=\mdisk/\Npl$. The system parameters are same as in \autoref{fig:planetesimal_dist_stages}. Different tracks of the same color denote evolution of different realisations, created using different random seeds, of the same  model system. In each of these cases the total mass of interacting planetesimals is roughly the same, $\minteract/\mplanet\sim0.1$. The overall evolution of planets do not depend on $\Npl$ (or $\mpl$) as long as $\Npl$ is sufficiently large, or equivalently, $\mpl/\mplanet$ is sufficiently small.}
\label{fig:diff_npl_evol}
\end{figure}

In \autoref{fig:diff_npl_evol} we show the evolution of the offset ($\epsilon$) for the system shown in \autoref{fig:evo_nr} for two different values of $\mpl$, keeping the total disk mass ($\mdisk/\mplanet = 0.5$) fixed. In the first case, we run four different realizations with $\Npl = 5000$ according to the criteria discussed in \autoref{sec:num2} (blue; each curve represents an independent realization with a different random seed). In the other set of simulations, we again run four different realizations with $\Npl = 10000$ (orange). The only difference between these two sets is that the individual planetesimal mass $\mpl$ is different by a factor of two from each other. The $\epsilon$ evolution in the two sets are very similar to each other. The differences between the simulations between the sets is well within the differences coming from statistical fluctuations within a set. This test indicates that as long as $\mpl$ is sufficiently small compared to $\mplanet$ (see discussion on \autoref{sec:num2}), the exact value of $\mpl$ does not effect the outcome of the simulations in a statistically significant manner. Hence, each planetesimal in our study can also be considered as a swarm of smaller bodies, interacting with the planets.

\clearpage

\bibliographystyle{aasjournal}

\end{document}